\documentclass[12pt]{article}
\usepackage{epsfig}
\usepackage{amsfonts}
\usepackage{graphicx}
\usepackage{amsmath}
\usepackage{latexsym}
\usepackage{amstext}
\usepackage{amssymb}
\usepackage{array}
\usepackage{multirow}
\usepackage{color}
\usepackage{afterpage}
\usepackage{epstopdf}
\usepackage{caption}
\usepackage[normalem]{ulem}

\usepackage{graphicx,epsfig}
\usepackage{amsmath}
\usepackage{latexsym}
\usepackage{amstext}
\usepackage{ulem}
\usepackage{amssymb}
\usepackage{array}
\usepackage{multirow}
\usepackage{tikz}
\usepackage{authblk}
\usepackage{cite}
\usepackage[margin=0.6in]{geometry}
\usepackage{url}

\usepackage[caption=false]{subfig}
\newcommand{\beq}{\begin{equation}}
\newcommand{\eeq}{\end{equation}}
\newcommand{\beqq}{\begin{equation*}}
\newcommand{\eeqq}{\end{equation*}}
\newcommand{\beqa}{\begin{eqnarray}}
\newcommand{\eeqa}{\end{eqnarray}}

\newcommand{\ba}{\bea \begin{array}}
	\newcommand{\ea}{\end{array} \eea}

\newcommand{\bc}{\begin{center}}
	\newcommand{\ec}{\end{center}}
\newcommand{\p}{\partial}

\newcommand{\ds}{\displaystyle}

\newcommand{\s}{\mbox{\boldmath$s$}}
\newcommand{\sigmab}{\mbox{\boldmath$\sigma$}}

\newcommand{\B}{\mbox{\boldmath$B$}}

\begin{document}
\title{Modeling bursting in neuronal networks using facilitation-depression and afterhyperpolarization}
\author{Lou Zonca$^{*,\dagger}$ and David Holcman\footnote{Group of Computational Biology and Applied Mathematics, Institut de Biologie de l'\'Ecole Normale Sup\'erieure, 46 rue d'Ulm 75005 Paris, France.\\ $\dagger$ Sorbonne University, Pierre et Marie Curie Campus, 5 place Jussieu 75005 Paris, France.}}
\date{August 31, 2020}
\maketitle
\begin{abstract}
In the absence of inhibition, excitatory neuronal networks can alternate between bursts and interburst intervals (IBI), with heterogeneous length distributions. As this dynamic remains unclear, especially the durations of each epoch, we develop here a bursting model based on synaptic depression and facilitation that also accounts for afterhyperpolarization (AHP), which is a key component of IBI. The framework is a novel stochastic three dimensional dynamical system perturbed by noise: numerical simulations can reproduce a succession of bursts and interbursts. Each phase corresponds to an exploration of a fraction of the phase-space, which contains three critical points (one attractor and two saddles) separated by a two-dimensional stable manifold $\Sigma$. We show here that bursting is defined by long deterministic excursions away from the attractor, while IBI corresponds to escape induced by random fluctuations. We show that the variability in the burst durations, depends on the distribution of exit points located on $\Sigma$ that we compute using WKB and the method of characteristics. Finally, to better characterize the role of several parameters such as the network connectivity or the AHP time scale, we compute analytically the mean burst and AHP durations in a linear approximation. To conclude the distribution of bursting and IBI could result from synaptic dynamics modulated by AHP.
\end{abstract}
\section*{Introduction}
Neuronal networks can exhibit periods of synchronous high-frequency activity called bursts separated by interburst intervals (IBI), corresponding to low amplitude time periods. Bursting can either be due to intrinsic channel activities driven by calcium and voltage-gated channels or by collective synchronization of large ensemble of neuronal cells \cite{scholarpediaBursts}. Yet the large distributions observed in electrophysiological recordings of bursting and IBI remains unclear.\\
Bursting is a fundamental feature of Central Pattern Generators such as the respiratory rhythm in the pre-Bötzinger complex \cite{Feldman1991,Feldman2016}, mastication or oscillatory motor neurons \cite{Marder2005} which are involved in the genesis and maintenance of rhythmic patterns. Interestingly, several coupled pacemaker neurons receiving an excitatory input from tonic firing neurons can either lead to bursting, tonic spiking or resting depending on the values of the channel conductances and the neuronal coupling level \cite{Butera1999_1,Butera1999_2,Delnegro2001}.\\
Bursts that emerge as a network property have been studied using different modeling approaches such as coupled integrate and fire neurons \cite{brunelhakim1999,Hansel2001}, improved recently by adding noise to connected Hodgkin-Huxley type neurons, to allow desynchronisation \cite{Desroches2019}. Bursting can also depend on the balance between excitatory and inhibitory neurons: coupling excitatory neurons results in in-phase bursting within the network, whereas inhibitory coupling leads to anti-phase dynamics \cite{Shi2009}. Furthermore, time-delays \cite{HanselPRL2005} play a crucial role in synchronisation, by generating coherent bursting, specifically when the time-delays are inversely proportional to the coupling strength \cite{Liang2009}.\\
Rhythm generation based on network bursting also depends on the bursting frequency and the interburst intervals. Synaptic properties shape the genesis and maintenance of bursts \cite{Staley, Verderio_Bacci, Cohen_Segal2009}. Synaptic short-term plasticity modeled in the mean-field approximation, is based on facilitation, depression and network firing rate \cite{Tsodyks1997}. Long interburst intervals have been generated by introducing a two state synaptic depression \cite{Guerrier2015}. Interestingly, different levels of facilitation and depression lead to various network dynamics \cite{Barak2007} such as resting, bursting, spiking and, when noise is added, to Up and Down state transitions \cite{Holcman_Tsodyks2006}. Such models were used to interpret bursting in small hippocampal neuronal islands \cite{DaoDuc2015} to show that the correlation between successive bursts could result from synchronous depressing-facilitating synapses.\\
However in all these models, the distribution of Bursting and IBI durations remains unclear. In particular,  the IBI in hippocampal pyramidal neurons is shaped by various type of potassium and calcium ionic channels \cite{AHP_review,deSevilla_AHP,Tzingounis2007,Tzingounis2008}, leading to medium and slow hyperpolarizing currents in the cells, a phenomenon known as afterhyperpolarization (AHP). AHP results from the activation of these slow and fast potassium channels, but their exact biophysical properties and distributions are not fully known. Thus, we decided here to model the consequences of these channels by using a phenomenological approach to reproduce the shape of the AHP. We analyse this model using WKB methods to determine how the distribution of bursts durations  depends on some properties of exit points in the phase-space. Furthermore, we wish to better understand how the IBI durations depend on various parameters such as the network connectivity, the AHP time scales and the facilitation-depression dynamics. We do so by deriving analytical formulas for the burst and AHP durations using a linear approximation of our model. \\
The manuscript is organized in two main sections: in the first one, we introduce a  phenomenological three-dimensional dynamical system, where we have added the effect of AHP to the facilitation-depression model by modifying the dynamics in certain portion of the phase-space. Noise perturbation on the voltage variable can produce bursting periods followed by IBI. We describe the phase-space that contains three critical points (one attractor and two saddles). Moreover, we relate the distribution of burst durations to the one of the exit points on the stable manifold, delimiting the region of non bursting trajectories of the stable equilibrium. In the second section, we use a linear approximation of the phenomenological system we introduced in section 1, to obtain a closed relation between the burst and AHP durations and key parameters. Finally, we study how the network connectivity, facilitation and depression parameters influence the burst and IBIs.
\section{A facilitation-depression model with AHP}
\subsection{Model description} \label{ModelDescription}
\begin{figure} \centering
	\includegraphics[scale=0.8]{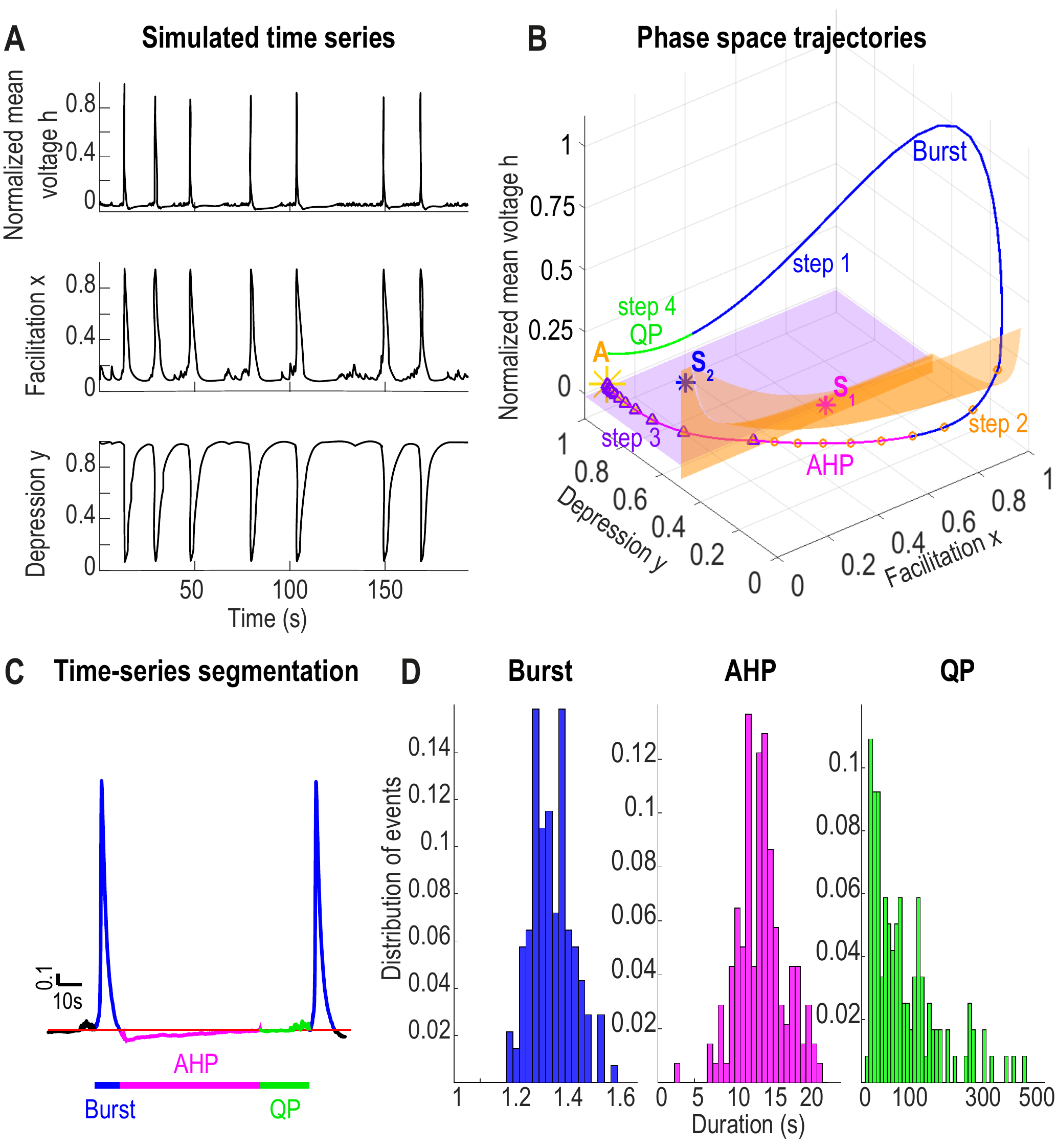}
	\caption{\textbf{Depression-facilition-AHP model}. \textbf{A}. Voltage time series (parameter $h$) normalized, the facilitation $x$ and the depression $y$ (lower) simulated from eq. \eqref{AHP_model}. \textbf{B}. Three dimensional phase-space showing a trajectory, decomposed into a QP (green), a burst (blue) and an AHP (pink) phase. The phase-space is divided into 3 regions: 1) the medium dynamics (step 2) of hyperpolarization where $\tau_0=\tau_{mAHP}, \text{ \& } T_0 = T_{AHP}$ under and right of the orange surface. In this region the trajectory is highlighted with orange circles. 2) The slow recovery dynamics (step 3, $\tau_0=\tau_{sAHP} \text{ \& } T_0 = 0$, region under the purple plan), where the trajectory is highlighted with purple triangles. 3) The fast dynamics (steps 1 and 4, $\tau_0=\tau \text{ \& } T_0 = 0$). Here $Y_{AHP} = 0.85$, $Y_h =0.5$ and $H_{AHP}=-7.5$. \textbf{C}. Segmentation of the time series in burst (blue) and IBI (AHP, pink) and QP (green). \textbf{D}. Distribution of duration for bursting (left, blue), AHP (center, pink) and QP (right, green) for numerical simulations lasting $10^4s$.}  \label{FigureModel}
\end{figure}

Since AHP involves the combination of several types of slow and fast potassium channels, to avoid entering into a difficult choice of channels, we decided instead to use a coarse-grained representation. We thus rather model the consequences of channel activity by modifying the facilitation-depression short-term synaptic plasticity model. This is well accounted for by a mean-field system of equations for a sufficiently well connected ensemble of neurons. The stochastic dynamical system consists of three equations \cite{Tsodyks1997, DaoDuc2015} for the mean voltage $h$, the depression $y$, and the synaptic facilitation $x$:
\beqa \label{sys}
\tau \dot{h} &=& - h + Jxy h^+ +\sqrt{\tau}\sigma \dot{\omega}\nonumber\\
\dot{x} &=& \dfrac{X-x}{t_f} + K(1-x) h^+ \\
\dot{y} &=& \dfrac{1-y}{t_r} - L xy h^+ , \nonumber
\eeqa
The population average firing rate is given by $h^+ = max(h,0) $, which is a linear threshold function of the synaptic current \cite{Holcman_Tsodyks2006}. The term $Jxy$ reflects the combined effect of synaptic short-term dynamics on the network activity. The second equation describes facilitation, while the third one describes depression. The mean number of connections (synapses) per neurons is accounted for by the parameter $J$ \cite{Bart}. We previously distinguished \cite{DaoDuc2015} the parameters $K$ and $L$ which describe how the firing rate is transformed into molecular events that are changing the duration and the probability of vesicular release respectively. The time scales $t_f$ and $t_r$ define the recovery of a synapse from the network activity. Finally, $\dot \omega$ is an additive Gaussian noise and $\sigma$ its amplitude, it represents fluctuations in the firing rate. \\
The model \eqref{sys} does not account for long AHP periods, where the voltage is hyperpolarized and then slowly depolarized due to potassium channels \cite{AHP_review}, leading to a refractory period that is not accounted for in the facilitation-depression model. To account for AHP, we thus incorporated changes in the facilitation-depression model by introducing two features: 1) a new equilibrium state representing hyperpolarization, after the peak response 2) a slow recovery with two timescales (medium and slow) to describe the slow transient to the steady state. The new equations are
\beq
\arraycolsep=1.4pt\def\arraystretch{2.0}
\begin{array}{r c l}
	\tau_0 \dot{h}&=&
	-(h-T_0) + Jxy(h-T_0)^+ +\sqrt{\tau_0}\sigma \dot \omega\\
	\dot{x} &=&\cfrac{X-x}{\tau_f} + K(1-x)(h-T_0)^+ \\
	\dot{y} &=&\cfrac{1-y}{\tau_r} - Lxy(h-T_0)^+.\\
\end{array}
\label{AHP_model}
\eeq
These changes lead to a piece-wise system that decomposes into four steps:
\begin{itemize}
	\item[-] {\bf step 1: burst phase.} It is defined when the dynamics fall into the subspace $\{y>Y_{AHP}$ and $h \geq H_{AHP}$ (fig. \ref{FigureModel}B purple surface)$\}$. During this phase the time constant $\tau_0$ of $h$ is fixed to $\tau_0=\tau$ and the resting value of $h$ is $T_0=T$ (see Table \ref{tableParam}). 
	\item[-] {\bf Step 2: depression phase.} In this phase, the depression parameter $y$ increases ($\dot{y}>0 \iff y<\cfrac{1}{1+Lx(h-T_0)}$, fig. \ref{FigureModel}B curved orange surface), and it lasts until $y$ reaches the threshold $Y_h$ (i.e. $y<Y_h$, fig. \ref{FigureModel}B vertical orange surface). During this phase the parameters are $\tau_0=\tau_{mAHP}$ and $T_0=T_{AHP}<T$. These parameter values forces the voltage to hyperpolarize.
	\item[-] {\bf Step 3: return to steady state.} In that phase,  the depression $y$ is still increasing ($\dot{y}>0$), with the condition that $Y_{AHP}<y$ or $h < H_{AHP}$. During this phase, we change the time constant to $\tau_0=\tau_{sAHP}$ and the resting value of $h$ is set to its initial value $T_0=T$. These modifications accounts for the slow recovery from hyperpolarization to the resting state, this phase ends when $y$ reaches the second threshold $Y_{AHP}$ and $h$ reaches its threshold $H_{AHP}$.
	\item[-]{\bf Step 4: resting state.} This phase models the fluctuations of the voltage around the steady state due to noise. The conditions and parameters are the ones of step 1 ($\{y>Y_{AHP}$ and $h \geq H_{AHP}\}$, $\tau_0=\tau$ and $T_0=T$).
\end{itemize}
The values of the parameters for the classical facilitati,on-depression part are chosen in agreement with \cite{Tsodyks1997,DaoDuc2015,Holcman_Tsodyks2006,Barak2007}, while the AHP parameters ($T_{AHP}, \tau_{mAHP} \text{ and }\tau_{sAHP}$, Table \ref{tableParam}) are consistent with the biological observations \cite{AHP_review}.\\
Numerical simulations of equations \eqref{AHP_model} with a sufficient level of noise exhibit spontaneous bursts in the voltage variable followed by AHP periods (fig. \ref{FigureModel}A-B). \\
We segmented the simulated time series into two phases: bursting (fig. \ref{FigureModel}C, blue) and IBI, which is further decomposed into an AHP period (pink) and a quiescent phase (QP) in  green. The quiescent phase is a period where the voltage fluctuates around its equilibrium value $h=0$. This segmentation allows us to obtain the distributions of burst, AHP and QP durations (fig. \ref{FigureModel}D).
\begin{center}
\begin{tabular}{l l l}
& Parameters & Values \\
\hline
$\tau$ & Fast time constant for $h$ & 0.05s \cite{Holcman_Tsodyks2006} \\
$\tau_{mAHP}$ & Medium time constants for $h$ & 0.15s \\
$\tau_{sAHP}$ & Slow time constants for $h$ & 5s \\
$J$ & Synaptic connectivity & 4.21 (\textit{modified: 3-5 in} \cite{Barak2007})\\
$K$ & Facilitation rate & 0.037Hz (\textit{modified: 0.04Hz in} \cite{DaoDuc2015})\\
$X$ & Facilitation resting value & 0.08825 (\textit{modified: 0.5-0.1 in} \cite{Barak2007})\\
$L$ & Depression rate & 0.028Hz (\textit{modified: 0.037Hz in} \cite{DaoDuc2015})\\
$\tau_r$ & Depression time rate & 2.9s (\textit{modified: 2-20s in} \cite{DaoDuc2015})\\
$\tau_f$ & Facilitation time rate & 0.9s (\textit{modified: 1.3s in} \cite{DaoDuc2015})\\
$T$ & Depolarization parameter & 0 \\
$\sigma$ & Noise amplitude & 3\\
$T_{AHP}$ & Undershoot threshold & -30\\
\hline
\end{tabular}
\end{center}\captionof{table}{Model parameters}\label{tableParam}
\subsection{Phase-space analysis}
Studying the phase-space of the deterministic system \eqref{AHP_model} is a key step to analyze the stochastic dynamics.
\subsubsection{Equilibrium points}
\begin{figure} \centering
	\includegraphics[scale=0.8]{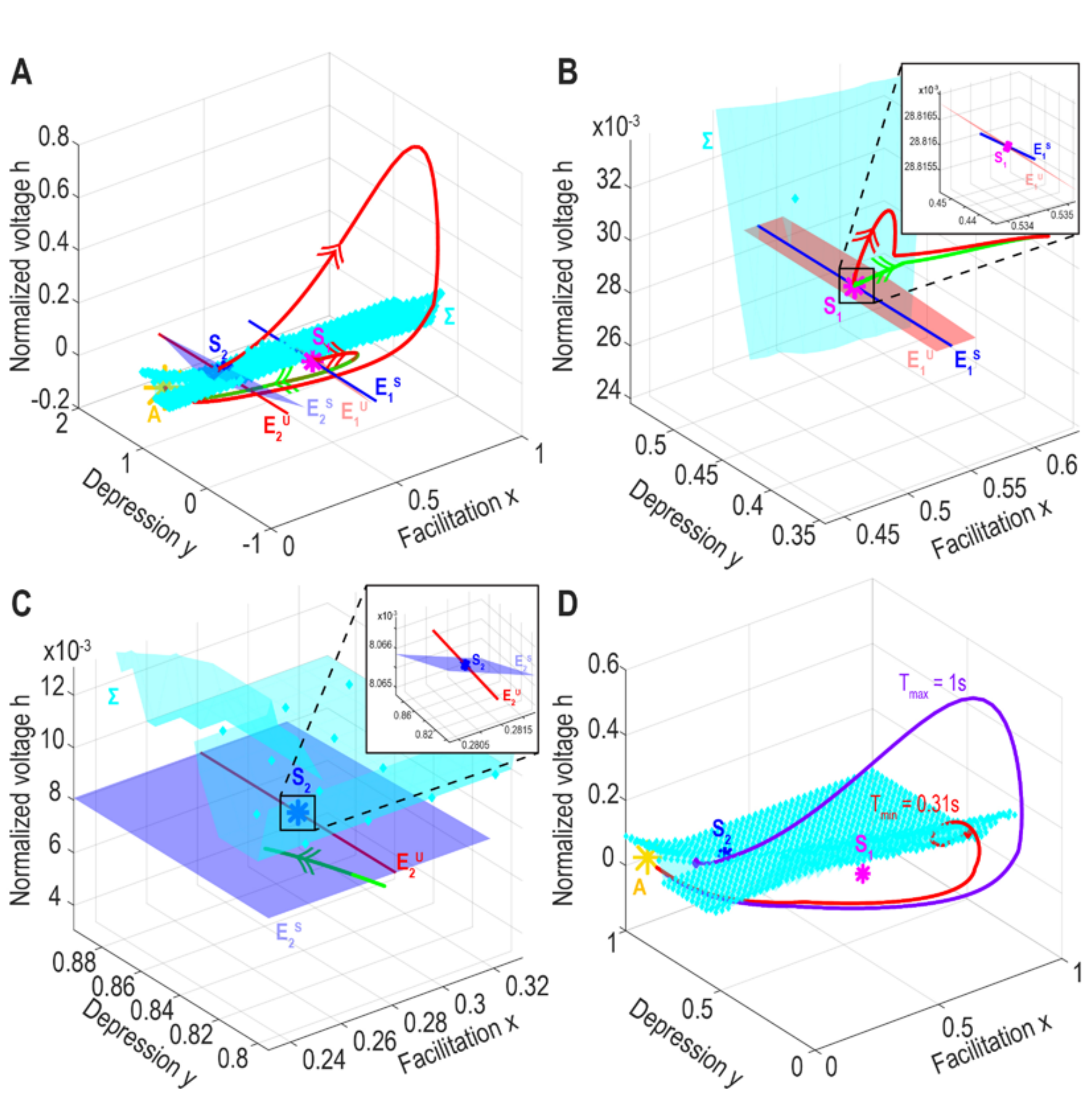} \caption{\textbf{ Phase-space dynamics starting from saddle points.} \textbf{A}. Repulsive trajectories starting near $S_1$ (pink) and $S_2$ (blue) with corresponding stable and unstable eigenspaces $E^U$ and $E^S$. \textbf{B}. Inset around $S_1$. Linear 2D-unstable eigenspace $E^U_1$ (light red plane) and linear 1D-stable eigenspace $E^S_1$ (blue line). \textbf{C}. Magnification around $S_2$: linear 1D-unstable eigenspace $E^U_2$ (red line) and linear 2D-stable eigenspace $E^S_1$ (blue plane). \textbf{D}. Plot of the longest (purple) and shortest (red) trajectories starting just above $\Sigma$.}  \label{PhaseSpace}
\end{figure}

We first search for the equilibrium points. There are three of them:
\paragraph{Attractor.} The first equilibrium point $A$ is given by $h=0, x=X, y=1$ and the Jacobian at this point is given by
\beq\label{jac_A}
J_A = \arraycolsep=1.4pt\def\arraystretch{2.0}
\left(
\begin{array}{c c c}
	\cfrac{-1+JX}{\tau} \phantom{1234}& 0 \phantom{1234}& 0 \\
	K(1-X) \phantom{1234}& - \cfrac{1}{\tau_f} \phantom{1234}& 0 \\
	LX \phantom{1234}& 0 \phantom{1234} & - \cfrac{1}{\tau_r}\\
\end{array}
\right).
\eeq
The eigenvalues  $(\lambda_1, \lambda_2, \lambda_3 ) = \left(\cfrac{-1+JX}{\tau}, - \cfrac{1}{\tau_f},- \cfrac{1}{\tau_r}\right)$ are real strictly negative (fig.\ref{FigureModel}B and \ref{PhaseSpace}A, yellow star). With the parameters of Table \ref{tableParam}, we obtain three orders of magnitude $|\lambda_1| = 12.6 \gg |\lambda_2| = 1.1 \gg |\lambda_3| = 0.34$. The dynamics near the attractor is thus very anisotropic, restricted to the plan perpendicular to the eigenvector associated to the highest eigenvalue $|\lambda_1|$.
\paragraph{Saddle-points $S_1$ and $S_2$.} The other steady-state solutions are given by $Jxy=1$ thus,
\beqq
\cfrac{X-x}{\tau_f}+K(1-x)(h-T-T_0)=0 \Leftrightarrow h=T+T_0+ \cfrac{x-X}{\tau_f K (1-x)},
\eeqq
leading to
\beqq
\cfrac{1-\cfrac{1}{Jx}}{\tau_r}-\cfrac{L}{J}\cfrac{X-x}{\tau_f K (1-x)}=0
\Leftrightarrow (J \tau_f K + L\tau_r)x^2 - (\tau_f K (J+1) + LX\tau_r)x + \tau_f K =0.
\eeqq
The discriminant is
\beq
\Delta = (\tau_f K (J+1) + LX\tau_r)^2 - 4(J \tau_f K + L\tau_r)\tau_f K > 0,
\eeq
leading to
\beq
\begin{array}{l}
	x_{1,2} = \cfrac{\tau_f K (J+1) + LX\tau_r \pm \sqrt{\Delta}}{2(J\tau_f K + L\tau_r)} \\
	y_{1,2}=\cfrac{1}{Jx_{1,2}}\\
	h_{1,2}=T+T_0+\cfrac{x_{1,2}-X}{\tau_f K (1-x_{1,2})}.\\
\end{array}
\eeq
The Jacobians at these points are
\beq
J_{S_{1,2}} =
\begin{pmatrix}
	0& \cfrac{Jy_{1,2}(h_{1,2}-T-T_0)^+}{\tau_0} & \cfrac{Jx_{1,2}(h_{1,2}-T-T_0)^+}{\tau_0} \\
	K(1-x_{1,2})& - \cfrac{1}{\tau_f}-K(h_{1,2}-T-T_0)^+ & 0 \\
	-\cfrac{L}{J} & -Ly_{1,2}(h_{1,2}-T-T_0)^+ & - \cfrac{1}{\tau_r}-Lx_{1,2}(h_{1,2}-T-T_0)^+.\\
\end{pmatrix}
\label{jac_saddle}
\eeq
With the parameter values of Table \ref{tableParam},  $y_{1,2}>Y_h$ and thus  $T_0=0$. Moreover, $\dot{y}_{|y_{1,2}}<0$ so $\tau_0=\tau$.\\
We computed numerically the eigenvalues of the matrices $J_{S_{1,2}}$.  The first saddle point $S_1$ has one real strictly negative eigenvalue and two complex-conjugate eigenvalues with positive real-parts $(\lambda_1,\lambda_2,\lambda_3) = (-5.06, 1.05 + 1.16i, 1.05 - 1.16i)$: $S_1$ is a saddle-focus (with a repulsive focus and a stable manifold of dimension 1, fig. \ref{PhaseSpace}B). The second saddle point $S_2$ has two real negative eigenvalues and one positive one
$(\lambda_1,\lambda_2,\lambda_3) = (-4.58, -0.25, 3.01)$, it is a saddle-point with a stable manifold of dimension two and unstable of dimension one (fig. \ref{PhaseSpace}C).
\subsubsection{Two dimensional separatrix $\Sigma$:  boundary of long excursions away from the stable equilibrium A}
\begin{figure}
\centering
\includegraphics[scale=0.8]{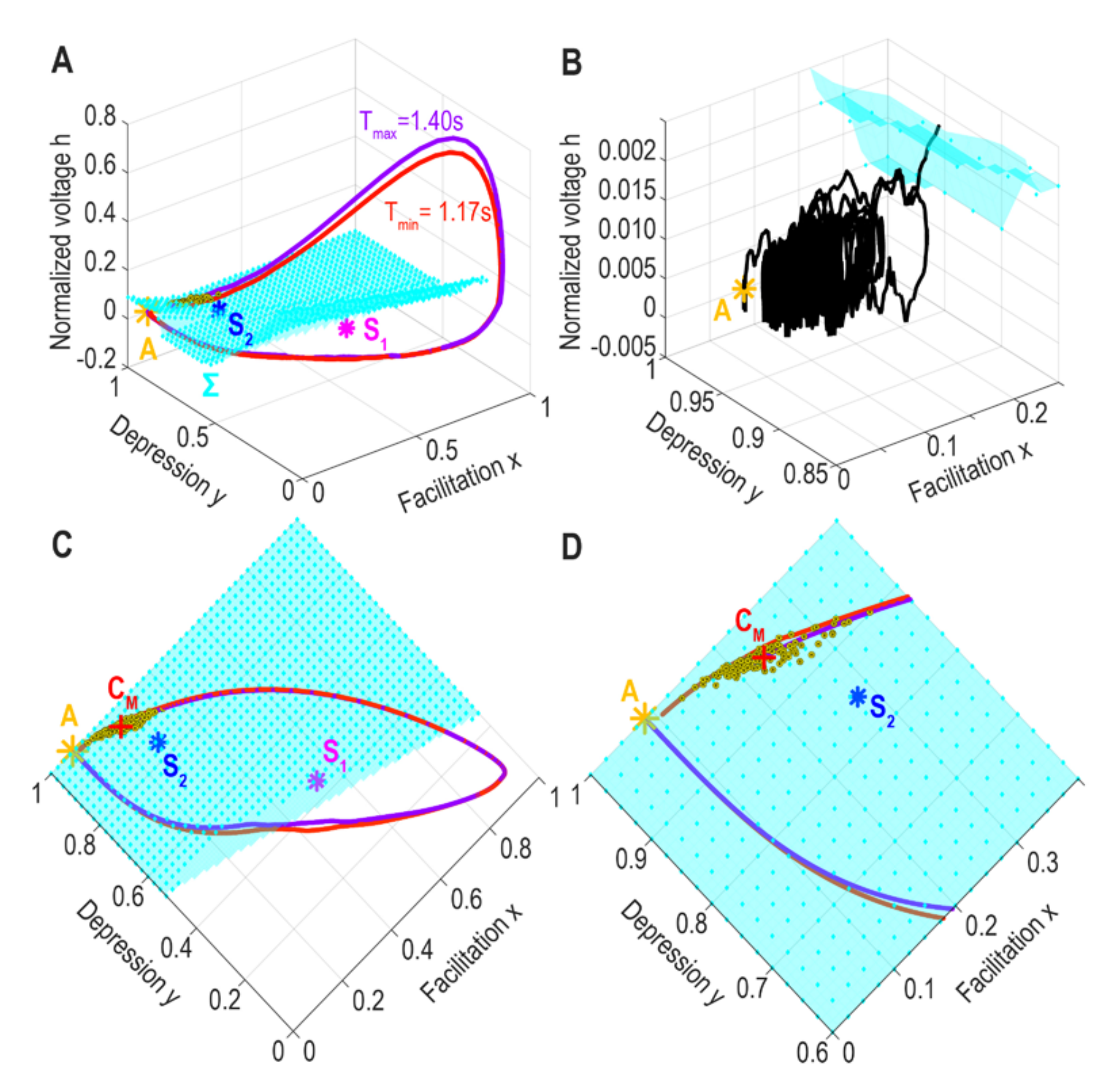}	\caption{\textbf{Stochastic dynamics in the phase-space of fig.\ref{PhaseSpace}.} \textbf{A}. Distribution of exit points (black dots on $\Sigma$) of system \eqref{AHP_model} starting from $A$ (5000s runs with variance $\sigma=3$) with the longest (purple) and shortest (red) trajectories. \textbf{B}. Example of an escaping trajectory (black) starting from $A$ (yellow star), making multiple short loops before exit. \textbf{C}. Top view of \textbf{A} and center of mass $C_M$ of the exit points (red cross). \textbf{D}. Magnification of  \textbf{C}.} \label{StocSysts}
\end{figure}

The deterministic trajectories can be compartmentalized in two categories: 1) Bursting trajectories, doing long excursions away from the attractor $A$ before going back and 2) trajectories going straightforwardly back to $A$. We determine the \textit{bursting boundary} as the separatrix surface $\Sigma$ (fig. \ref{PhaseSpace}, cyan surface) passing through $S_2$ (the stable manifold of $S_2$) and splitting the phase-space in 2 regions: $B_+$ situated "above" $\Sigma$ where deterministic trajectories define bursts and $B_-$, "below" $\Sigma$ where trajectories go straight back to $A$.\\
To determine $\Sigma$, we sampled the (h,x,y)-space with various initial conditions to determine the location where trajectories are confined to $B_-$ and other characterized by a long trajectory away from the attractor in $B_+$, which describes the bursting phase. Finally, we note that the shape of $\Sigma$ can become very complex away from the saddle-point $S_2$, however here our trajectories are confined within the square prism defined by $\{x \in [0,1] \text{ \& } y \in [0,1]\}$ and our numerical simulations show that in this domain the surface $\Sigma$ is still simple enough for our approximation and that it does split the phase-space in the two subdomains described (fig. \ref{PhaseSpace} and \ref{StocSysts} cyan surface).\\
We constructed the separatrix $\Sigma$ with a precision $\Delta h=0.01$ for a normalized amplitude of h to 1, which is smaller than the spatial scale of the stochastic component of the simulation $\sigma \sqrt{\tau \Delta t}\approx 0.07$.\\
To characterize the range of bursting durations, we further determined numerically the durations of the shortest (red) and longest (purple) trajectories, starting in the upper neighborhood of the separatrix $\Sigma$ and ending below $h=0$ (fig. \ref{PhaseSpace}D): we found that the fastest and shortest durations are 1s and 0.31s respectively. Note that these durations are measured after departure from $\Sigma$ (no return, which could be possible in the stochastic case), that could explain the difference with the burst duration histogram (Fig. \ref{FigureModel}D). The extreme trajectories are determined when we sampled the initial condition in the discretized approximation of $\Sigma$ by a grid $(x_k,y_q)= (k\Delta x, q\Delta y) \in [0,1]^2$, where we used the resolution $\Delta x=\Delta y=0.025$.
\subsection{Distribution of exit points} \label{exitPointsCalculations}
To characterize the distribution of bursting durations, we decided first to focus on the distribution of exit points from the region $B_-$ located on the surface $\Sigma$. Our rational was the dominant dynamics above $\Sigma$ is deterministic. Thus any fluctuation should come from the statistics of the exit points distribution.
\subsubsection{Distribution of exit points obtained from stochastic simulations} \label{exitPointsCalculationsSSimulations}
{We first ran stochastic simulations of system \eqref{AHP_model} with the attractor $A = (0,X,1)$ as initial point for a fixed noise amplitude. For each burst, we recorded the intersection point (exit point) of the trajectory and $\Sigma$ (fig. \ref{StocSysts}). In this region of the phase-space, the dynamics simplifies to the system \eqref{sys} without AHP, which can be written in the matrix form
\beq \label{matSyst}
\dot{\s} = \B(\s)+\sqrt{\sigmab} \dot{W}
\eeq
where $\s=(h,x,y)^T$ and
\beq \label{B}
\B(\s)= \begin{pmatrix}
b_1(\s) =-\cfrac{h}{\tau} + \cfrac{Jxyh^+}{\tau}\\
b_2(\s)=\cfrac{X-x}{\tau_f} + K(1-x)h^+ \\
b_3(\s)=\cfrac{1-y}{\tau_r} - Lxyh^+\\
\end{pmatrix}
\eeq
and $\sqrt{\sigmab}=diag\left(\sqrt{\cfrac{\sigma}{\tau}},0,0\right)$.
\subsubsection{Distribution of exit points obtained from solving the Fokker-Planck equation} \label{exitPointsCalculationsFFP}
At this stage, we decided to compare the empirical distribution with the probability density function $q(\s)$ obtained form
the steady-state renewal Fokker-Planck equation (FPE) \cite{Schuss2,OPT}, when the initial point is $A$. This density is obtained by conditioning on trajectories of the process \eqref{matSyst} that are absorbed on $\Sigma$. It is solution of
\beq\label{fpe}
-\cfrac{\partial}{\partial h}\left[\cfrac{(Jxy-1)h}{\tau}q\right]-\cfrac{\partial}{\partial x}\left[\left(\cfrac{X-x}{\tau_f}+K(1-x)h\right)q\right]	-\cfrac{\partial}{\partial y}\left[\left(\cfrac{1-y}{\tau_r}-Lxyh\right)q\right]+
\cfrac{\sigma}{2\tau}\cfrac{\partial^2}{\partial h^2}q = \delta(\s-A)
\eeq
where
\beq \label{boundCond}
q(\s | A)=0 \hbox{ for } \s \in \Sigma.
\eeq
To solve equation \eqref{fpe}, we search for a WKB approximation of the  solution in the form
\beq \label{WKB}
q(\s| A)=Q_\sigma(\s)e^{\ds- \cfrac{\psi (\s)}{\sigma}},
\eeq
where $Q_\sigma$ is a regular function with the formal expansion
\beq
Q_\sigma(\s)=\sum_{i=0}^\infty Q_i(\s)\sigma^i.
\eeq
The function $\psi$ satisfies the eikonal equation \cite{Schuss2,OPT}
\beq \label{eikonal}
\cfrac{(Jxy-1)h}{\tau}\cfrac{\partial \psi}{\partial h}+\left(\cfrac{X-x}{\tau_f}+K(1-x)h \right)\cfrac{\partial \psi}{\partial x}+\left(\cfrac{1-y}{\tau_r}-Lxyh\right)\cfrac{\partial \psi}{\partial y}+\cfrac{1}{2\tau}\left(\cfrac{\partial \psi}{\partial h}\right)^2=0.
\eeq
We use the method of characteristics to solve the eikonal equation. Setting
\beq
p = \nabla \psi = \begin{pmatrix}
p_1 \\ p_2 \\p_3
\end{pmatrix},
\eeq
and
\beq \label{EDP_charac}
F(\s,\psi,p) = b_1(\s)p_1+b_2(\s)p_2+b_3(\s)p_3+\cfrac{1}{2\tau}p_1^2,
\eeq
the characteristics are given by
\beq \label{charac}
\begin{array}{r c l}
\cfrac{dh}{dt}&=&F_{p_1}=b_1+\cfrac{1}{\tau}p_1 \\
\cfrac{dx}{dt}&=&F_{p_2}=b_2 \\
\cfrac{dy}{dt}&=&F_{p_3}=b_3,
\end{array}
\eeq
\beq
\begin{array}{r c l}
\cfrac{dp_1}{dt}&=&-F_h=-\cfrac{Jxy-1}{\tau}p_1-K(1-x)p_2+Lxyp_3\\
\cfrac{dp_2}{dt}&=&-F_x=-\cfrac{Jyh}{\tau}p_1+\left(\cfrac{1}{\tau_f}+Kh\right)p_2+Lyhp_3\\
\cfrac{dp_3}{dt}&=&-F_y=-\cfrac{Jxh}{\tau}p_1+\left(\cfrac{1}{\tau_r}+Lxh\right)p_3
\end{array}
\eeq
and
\beq \label{dpsi}
\cfrac{d \psi}{dt}=\cfrac{1}{2\tau}p_1^2.
\eeq
We solve \eqref{charac}-\eqref{dpsi} starting at the attractor $A$, however, this characteristic will be trapped at $A$. To avoid this difficulty, we follow the method proposed in \cite{OPT} p.165-170, and we start from points located in a neighborhood $V_A$ of $A$.  In $V_A$, the solution of the eikonal equation has a quadratic approximation
\beq \label{quadraPsi}
\psi(\s)=\cfrac{1}{2}\s^TR\s+ o(|\s|^2).
\eeq
To find the matrix $R$, we linearized the eikonal equation around the attractor $A$
\beq \label{eikonlaeq}
(J_A\s)^T \cdot \nabla \psi +\cfrac{1}{2 \tau}p_1^2=0,
\eeq
where $J_A$ is the Jacobian defined in \eqref{jac_A}. Due to the noise present in only one coordinate, this matrix equation  \eqref{eikonlaeq} does not have a unique solution. We shall use the one given by
\beq
\psi(\s) \approx (1-JX)h^2.
\eeq
We now follow the method of reconstruction \cite{OPT} by choosing the initial points on the contours $\psi(\s) =\delta= 0.05$, that is
\beq
h=\pm \sqrt{\frac{\delta}{1-JX}} \approx 0.28.
\eeq
We then computed the characteristics numerically (fig. \ref{charactericsFig}A-B). \\
The final step to determine the exit points distribution is to solve the transport equation \eqref{transport}
\beq \label{transport}
\begin{split}
	\cfrac{1-Jxy}{\tau}\left(h\cfrac{\partial Q_0}{\partial h} + Q_0 \right)+\left(\cfrac{1}{\tau_f}+Kh \right) Q_0 - \left(\cfrac{X-x}{\tau_f}+K(1-x)h\right)\cfrac{\partial Q_0}{\partial x}
	+ \left(\cfrac{1}{\tau_r}+Lxh \right) Q_0 \\ -\left(\cfrac{1-y}{\tau_r}-Lxyh\right)\cfrac{\partial Q_0}{\partial y}
	-\cfrac{1}{\tau}\cfrac{\partial Q_0}{\partial h}\cfrac{\partial \psi}{\partial h}-\cfrac{Q_0}{2\tau}\cfrac{\partial^2 \psi}{\partial h^2}=0.
\end{split}
\eeq
To find $Q_0$, we follow the method from \cite{OPT} p.172-175: we rewrite equation \eqref{transport}
\beq \label{transShort}
\B \cdot \nabla Q_0 + \cfrac{1}{\tau}\cfrac{\partial Q_0}{\partial h}\cfrac{\partial \psi}{\partial h} = -\left(\nabla \cdot \B + \cfrac{1}{2 \tau} \cfrac{\partial^2 \psi}{\partial h ^2}\right) Q_0
\eeq
where $\B$ is defined in \eqref{B}.
Along the characteristics, \eqref{transShort} is
\beq
\cfrac{dQ_0(s(t))}{dt} = \nabla Q_0(s(t)) \cdot  \cfrac{d\s(t)}{dt} = -\left(\nabla \cdot \B(s(t)) + \cfrac{1}{2 \tau} \cfrac{\partial^2 \psi(s(t))}{\partial h ^2}\right) Q_0(s(t)).
\eeq
Our goal is to compute $Q_0$ on the separatrix and for that purpose, we need to evaluate $\cfrac{\partial^2 \psi(s(t))}{\partial h ^2}$ by differentiating the characteristics equations \eqref{charac}-\eqref{dpsi} with respect to the initial point $\s_0 = \s(0)$. Setting
\beq \label{matrixR}
\s_j(t) = \cfrac{\partial \s(t)}{\partial \s_0^j}, \phantom{12345} p_j(t) = \cfrac{\partial p(t)}{\partial \s_0^j}, \phantom{12345}
\cfrac{\partial^2 \psi (\s(t))}{\partial s^i\partial s^j} =R^{i,j}(t),
\eeq
we have $R(t) = P(t)S(t)^{-1}$, where $P(t)$ (resp. $S(t)$) is the matrix with columns $p_j(t)$ (resp. $\s_j(t)$). The initial conditions are
\beq
s_j^i(0)=\delta_{i,j}, \phantom{12345} p_j^i(0)=\cfrac{\partial^2 \psi (0)}{\partial s^i\partial s^j} =R^{i,j}.
\eeq
The dynamic has the form
\beq
\arraycolsep=1.4pt\def\arraystretch{2.3}
\begin{array}{r c c c c l}
	\cfrac{d \s_1^1 }{dt}&=&\cfrac{dh_1}{dt}&=&\left(\cfrac{\partial b_1}{\partial h}+\cfrac{1}{\tau}\cfrac{\partial p^1}{\partial h}\right)h_1\\
	\cfrac{d \s_2^1 }{dt}&=&\cfrac{dh_2}{dt}&=&\left(\cfrac{\partial b_1}{\partial h}+\cfrac{1}{\tau}\cfrac{\partial p^1}{\partial h}\right)h_2\\
	\cfrac{d \s_3^1 }{dt}&=&\cfrac{dh_3}{dt}&=&\left(\cfrac{\partial b_1}{\partial h}+\cfrac{1}{\tau}\cfrac{\partial p^1}{\partial h}\right)h_3\\
	\cfrac{d \s_1^2 }{dt}&=&\cfrac{dx_1}{dt}&=&\cfrac{\partial b_2}{\partial x}x_1\\
	\cfrac{d \s_2^2 }{dt}&=&\cfrac{dx_2}{dt}&=&\cfrac{\partial b_2}{\partial x}x_2\\
	\cfrac{d \s_3^2 }{dt}&=&\cfrac{dx_3}{dt}&=&\cfrac{\partial b_2}{\partial x}x_3\\
	\cfrac{d \s_1^3 }{dt}&=&\cfrac{dy_1}{dt}&=&\cfrac{\partial b_3}{\partial y}y_1\\
	\cfrac{d \s_2^3 }{dt}&=&\cfrac{dy_2}{dt}&=&\cfrac{\partial b_3}{\partial y}y_2\\
	\cfrac{d \s_3^3}{dt}&=&\cfrac{dy_3}{dt}&=&\cfrac{\partial b_3}{\partial y}y_3\\
\end{array}
\eeq
and because we are only interested in $R^{1,1}$ we only need to compute the first row of $P(t)$, thus
\beq
\arraycolsep=1.4pt\def\arraystretch{2.3}
\begin{array}{r c c c c l}
	\cfrac{d p_1^1(t)}{dt}=\left(-\cfrac{Jxy-1}{\tau}\cfrac{\partial p^1}{\partial h}-K(1-x)\cfrac{\partial p^2}{\partial h}+Lxy \cfrac{\partial p^3}{\partial h}\right) h_1 \\
	\cfrac{d p_2^1(t)}{dt}=\left(-\cfrac{Jxy-1}{\tau}\cfrac{\partial p^1}{\partial h}-K(1-x)\cfrac{\partial p^2}{\partial h}+Lxy \cfrac{\partial p^3}{\partial h}\right) h_2\\
	\cfrac{d p_3^1(t)}{dt}=\left(-\cfrac{Jxy-1}{\tau}\cfrac{\partial p^1}{\partial h}-K(1-x)\cfrac{\partial p^2}{\partial h}+Lxy \cfrac{\partial p^3}{\partial h}\right) h_3.
\end{array}
\eeq
In the limit $t \rightarrow \infty$ the characteristic that hits the saddle point $S_2$ is tangent to the separatrix and $-\left(\nabla \cdot \B + \cfrac{1}{2 \tau} \cfrac{\partial^2 \psi}{\partial h ^2}\right) Q_0 \rightarrow -\nabla \cdot \B _{|S_2} \approx 1.82$. Indeed, $\cfrac{\partial^2 \psi}{\partial h ^2}$ tends to $0$ near the saddle point $S_2$ as shown in fig. \ref{charactericsFig}C. Thus, near the saddle point, we have
\beq
\cfrac{dQ_0(s(t))}{dt} =-(\nabla \cdot \B _{|S_2}+o(1))Q_0(s(t)).
\eeq
The solution is approximated by
\beq \label{Q0_1}
Q_0(s(t)) = Q_0(s(0))e^{\ds -\nabla \cdot \B _{|S_2}t(1+o(1))}.
\eeq
Finally, the characteristic $s(t)$ near the saddle point $S_2$ can be expressed with respect to the arc length $\tilde{s}$:
\beq
\tilde{s}(t) \approx  \int_0^t \sqrt{\dot{s_2}(u)^2}du,
\eeq
where  $s_2$ is the dominant coordinate of $s \in \Sigma$ in the eigenvectors basis of the jacobian $J_{S_2}$ of system \eqref{AHP_model} at $S_2$,  ($\lambda_1 \approx -4.58$ and $\lambda_2 \approx -0.25$), thus locally
\beq
\tilde{s}(t) \approx \int_0^t \sqrt{s_2(0)e^{2\lambda_2u}}du,
\eeq
and
\beq \label{stilde}
\tilde{s}(t) \approx  \int_0^t \sqrt{s_2(0)e^{2\lambda_2u}}du = s_2(0) \cfrac{e^{\lambda_2t}-1}{\lambda_2}.
\eeq
Finally, using \eqref{Q0_1} and \eqref{stilde}, we obtain locally
\beq
Q_0(\tilde{s}) = Q_0(0)\tilde{s}^{-\cfrac{\nabla \cdot \B _{|S_2}}{\lambda_2}},
\eeq
where $ -\cfrac{\nabla \cdot \B _{|S_2}}{\lambda_2}\approx-7.23$.\\
The distribution of exit points is constructed from the solution $q$ of the FPE \eqref{fpe} by accounting for the boundary layer function $q_{\sigma}$ that has to be added to the transport solution in the form $Q_0q_\sigma$, such that  this product now satisfies the absorbing boundary condition \eqref{boundCond}. We do not compute here $q_\sigma$ as the computation follows the one of \cite{OPT} p. 182-183 near the separatrix. It is a regular function of the form $-\sqrt{\cfrac{2}{\pi}}\ds \int_0^{\rho \gamma(s_1,s_2)/\sqrt{\sigma}} e^{\ds -\eta^2/2} d\eta$, where $\rho$ is the distance to the separatrix $\Sigma$ in a neighborhood of $S_2$ and $\gamma(s_1,s_2)$ a regular function.\\
Finally, the exit point distribution per unit surface $d\s$ is given by
\beq \label{sol}
p_\Sigma(\tilde{s}|\s_0)=\cfrac{J(\tilde{s}|\s_0)\cdot\mathbf{\nu}(\tilde{s})d\tilde{s}}{{\oint_\Sigma  J(\tilde{s}|\s_0) \cdot \nu(\tilde{s}) d\tilde{s}}} \hbox{ for } \tilde{s} \in \Sigma
\eeq
where the probability flux is
\beq \label{flux}
J(\tilde{s}|\s_0)=\begin{pmatrix}
	\cfrac{Jxy-1}{\tau}hq(\tilde{s}) - \cfrac{\sigma}{2\tau}\cfrac{\partial q(\tilde{s})}{\partial h} \\
	\left(\cfrac{X-x}{\tau_f}+K(1-x)h\right)q(\tilde{s}) \\
	\left(\cfrac{1-y}{\tau_r}-Lxyh\right)q(\tilde{s})
\end{pmatrix},
\eeq
and $\nu(\tilde{s})$ is the unit normal vector at the point $\tilde{s}$. The flux is computed by differentiating expression \eqref{WKB} on the boundary
\beq
q(\tilde{s}|\s_0)=q_{\sigma}(\tilde{s}) Q_0(\tilde{s})e^{\ds- \cfrac{\psi (\tilde{s})}{\sigma}}.
\eeq
We obtain
\beq \label{fluxFinal}
\arraycolsep=1.6pt\def\arraystretch{3}
\begin{array}{r c l}
J(\tilde{s}|\s_0)\cdot\mathbf{\nu}(\tilde{s})d\tilde{s}&=&-\sqrt{\cfrac{2\sigma}{\pi}}q(\tilde{s}|\s_0) \gamma(s_1,s_2)d\tilde{s}\\
&=& K_0\tilde{s}^{-\cfrac{\nabla \cdot \B _{|S_2}}{\lambda_2}}e^{\ds- \cfrac{\psi (\tilde{s})}{\sigma}}d\tilde{s},
\end{array}
\eeq
where $\gamma(s_1,s_2)$ has been approximated by its value at $\tilde{s}=0$. Furthermore, in the limit $\tilde{s} \rightarrow 0$, $\tilde{s}^{-\cfrac{\nabla \cdot \B_{|S_2}}{\lambda_2}}$ tends to infinity, however it is compensated by $e^{\ds- \cfrac{\psi (\tilde{s})}{\sigma}}$ which is small enough, as we observe numerically. We plotted the distribution of exit points in fig. \ref{charactericsFig}D-E for $K_0=1$. Finally, we compare the distribution $p_\Sigma$ with the one obtained from the stochastic simulations of system \eqref{AHP_model} with the same level of noise ($\sigma =3$). Both distributions are peaked, showing that the exit points are constrained in a small area of the separatrix.\\
To conclude this part, our two different numerical methods confirm that the exit point distribution is peaked, thus the trajectories associated to the bursting periods are confined in a tubular neighborhood of a generic trajectory and thus the distribution of the bursting times is peaked, as observed in fig. \ref{FigureModel}D. Finally, the distribution of bursting durations should be quite concentrated near a deterministic value.
\begin{figure}
	\includegraphics[scale=0.8]{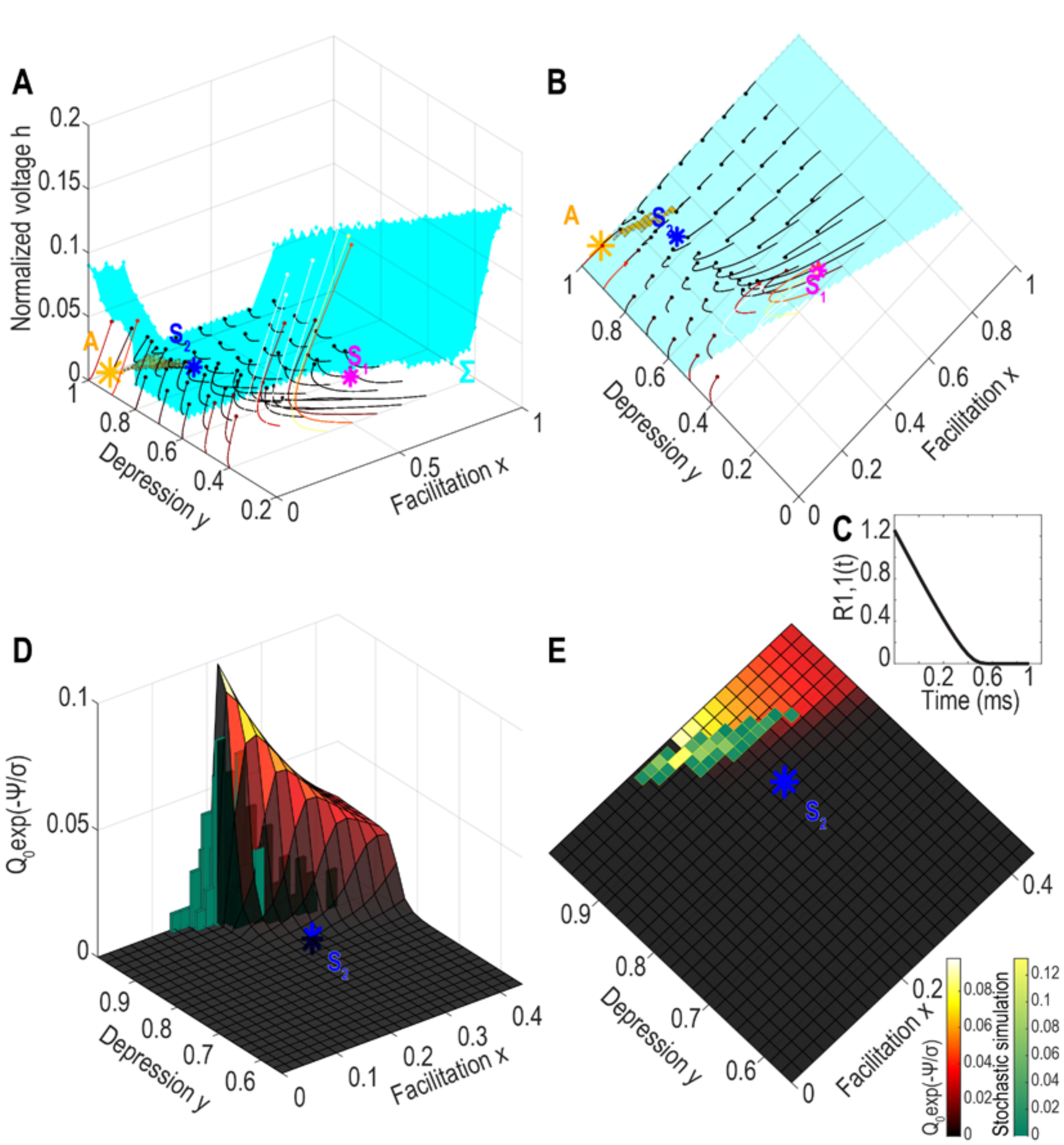}
	\caption{\textbf{Exit point distributions: characteristics vs stochastic realizations.} \textbf{A-B}. Characteristics crossing the separatrix $\Sigma$ (the darker the line color is, the lower the value of $\psi$ on $\Sigma$ is) and distribution of exit points obtained from numerical simulations (yellow); visualized with two different angles. \textbf{C}. Element $R^{1,1}(t) = \cfrac{\partial^2 \psi}{\partial h ^2}$ of the matrix \ref{matrixR} vs time along the characteristics computed numerically. \textbf{D-E}. PDF of the exit points $p_{\Sigma}=Q_0e^{-\cfrac{\psi}{\sigma}}$ on the separatrix $\Sigma$ compared to the distribution obtained from the stochastic simulations (green histogram), visualized with two different angles.}\label{charactericsFig}
\end{figure}
\section{Computing the burst and AHP durations}
How the burst duration depends on specific parameters such as the neuronal connectivity, the time characteristics of depression, facilitation or AHP durations is usually very difficult to address from a computational point of view. Indeed, it requires integrating a nonlinear dynamical system. Most of the time, the sensitivity analysis to parameters is explored numerically by sampling a certain fraction of the phase space. However, we shall show here that it is possible to get some expressions for the bursting and IBI durations.
\subsection{Deriving explicit expressions for the bursting and the AHP durations}
In this section we develop an approximation procedure to compute the mean bursting and afterhyperpolarization durations from the AHP facilitation-depression model \eqref{AHP_model}.\\
The approximation procedure is based on the following considerations: because
in the first phases of burst and AHP, the voltage $h$ evolves much faster than the facilitation $x$ and depression $y$, to compute the duration of the bursting phase, we will replace the dynamics of $h$ in the depression and facilitation equations by a piecewise constant function $H(t)$ (fig.\ref{Hcreneau}). This approximation decouples the system \eqref{AHP_model}, thus $x$ and $y$ can be computed. Indeed, $h$ increases quickly from 0 to a high value in less than 100 ms and then decays. Since we are interested in the decay phase, we will freeze the value of $h$ in equation \eqref{simplifiedSyst} for $x$ and $y$ in the time interval $[0,t_1]$ (the time $t_1$ will be estimated in section \ref{t1andt2}) to a high value $H_1$ (to be determined).  Moreover, in the interval $[t_1, t_2]$, we will fix the value of $h$ to a constant $H_2$ (to be determined) to account for the AHP phase. Then we will re-compute $h$ using the approximated equations for $x$ and $y$. We will then examine numerically how the unperturbed and perturbed solutions differs. Note that we are interested in the longer time phase (thus the different behavior of the solutions in the short-time will not matter much) so that we will able to use the analytical formulas to compute the burst and the AHP durations.
\\
We shall now specify the function $H(t)$. In the bursting phase, it is constant equal to $H_1$ for $t \in [0;t_1]$. In the hyperpolarization phase, $H(t)=H_2$ for $t \in [0;t_2]$. For $t>t_2$ (that will also be specified in section \ref{t1andt2}), we choose $H(t)=0$ to account for the recovery phase.
\beq \label{Hvalues}
H(t) = \left\{\begin{array}{l l}
	H_1, & \text{for } t \in [0, t_1]\\
	H_2, & \text{for } t \in ]t_1, t_2]\\
	0 &\text{for } t > t_2.
\end{array} \right.
\eeq
The approximated system of equations becomes:
\beq
\arraycolsep=1.4pt\def\arraystretch{2.0}
\begin{array}{*3{>{\ds}}{r c l}}
	\tau_0 \dot{h}&=&-(h-T_0(t)) + Jxy(h-T_0(t))^+\\
	\dot{x} &=&\cfrac{X-x}{\tau_f} + K(1-x)H(t)\\
	\dot{y} &=&\cfrac{1-y}{\tau_r} - LxyH(t)\\
\end{array}
\label{simplifiedSyst}
\eeq
where the AHP is accounted for by changing the threshold and timescales as follows
\beq \label{T0values}
\begin{array}{c c c}
	T_0(t) = \left\{\begin{array}{l l}
		0 &\text{for } t \in [0, t_1]\\
		T_{AHP} &\text{for } t \in ]t_1, t_2]\\
		0 &\text{for } t>t_2
	\end{array} \right. & \text{ and }& \tau_0(t) = \left\{\begin{array}{l l}
		\tau &\text{for } t \in [0, t_1]\\
		\tau_{mAHP} &\text{for } t \in ]t_1, t_2]\\
		\tau_{sAHP} &\text{for } t>t_2.
	\end{array} \right.
\end{array}
\eeq
\begin{figure}[http!] \center
	\includegraphics[scale=0.8]{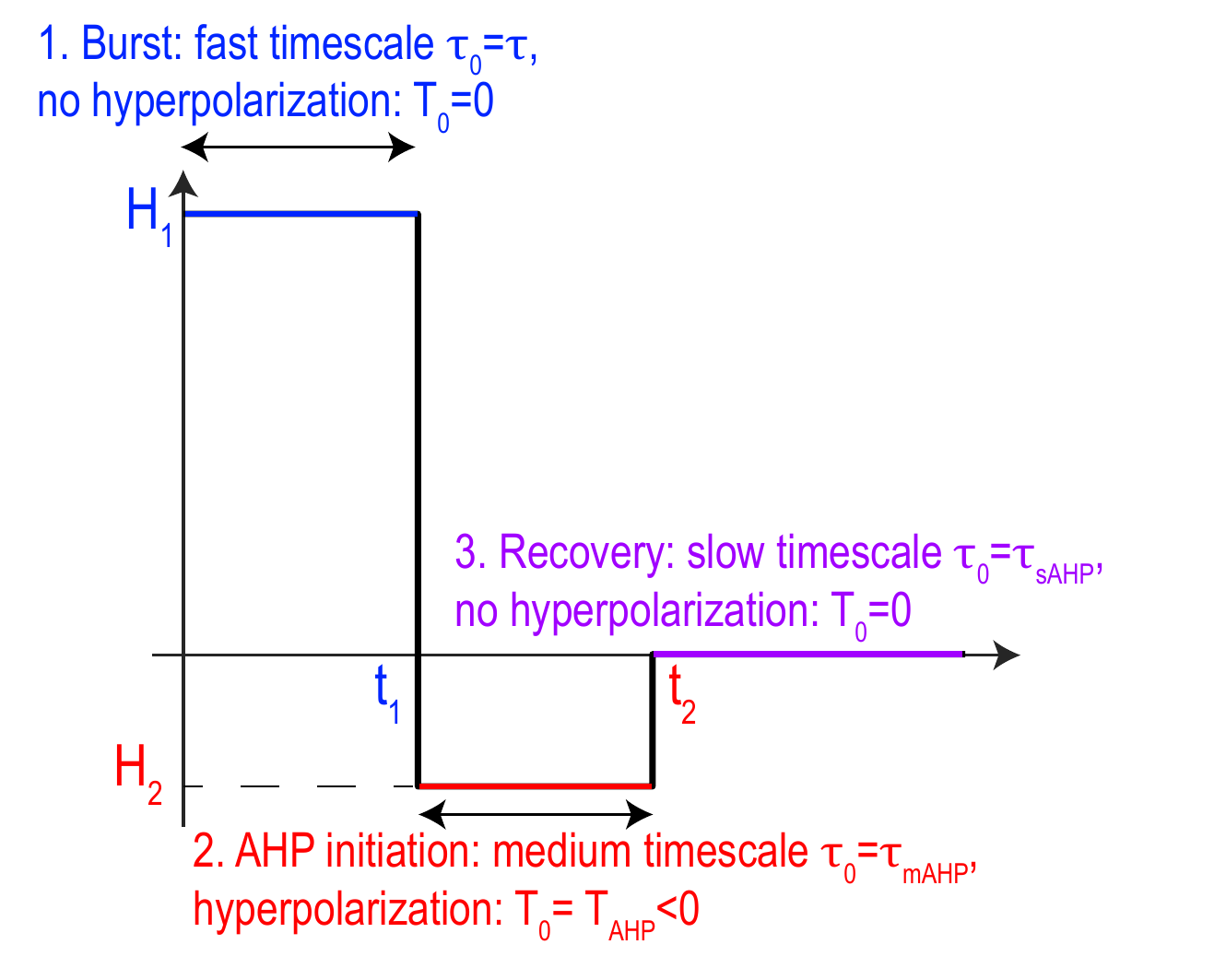}
	\caption{Approximated voltage step function $H(t)$.} \label{Hcreneau}
\end{figure}
\subsection{Explicit representation of the facilitation, depression and voltage variables in three phases} \label{xySolution}
\subsubsection*{Phase 1 $[0,t_1]$}
To integrate the facilitation and depression equations in \eqref{simplifiedSyst}, we note that during the bursting phase (fig. \ref{Hcreneau}, phase 1, blue) $H(t)=H_1$ with the initial conditions: $x(0)=X \text{ and } y(0)=1$ (resting values). We obtain
\beq
x(t)=A_1 e^{-\alpha_1 t}+B_1,  \label{x1approx}
\eeq
where
\beq
\alpha_1=\cfrac{1}{\tau_f}+KH_1, \phantom{123456} A_1=\cfrac{KH_1(X-1)}{\alpha_1},\phantom{123456} B_1=\cfrac{\cfrac{X}{\tau_f}+KH_1}{\alpha_1}.
\eeq
Injecting expression \eqref{x1approx} in the third equation of system \eqref{simplifiedSyst}, we obtain
\beqq
\dot{y}=\cfrac{1-y}{\tau_r}-L(A_1e^{-\alpha_1 t}+B_1)H_1y
\eeqq
The solution is
\beq
y(t)=\left(C_1+\frac{1}{\tau_r}\int_{0}^{t}\exp(f_1(s))ds\right)\exp(-f_1(t)), \label{y1exact}
\eeq
where the function
\beqq
f_1(t)=\beta_1t-\cfrac{LA_1H_1}{\alpha_1} e^{-\alpha_1 t}.
\eeqq
To approximate the integral $\int_0^t \exp(f_1(s))ds$, we use that $f_1$ is monotonic on the interval $[0;t_1]$, thus using a Taylor's expansion at order 1, we get
\beq
\int_{0}^{t} \exp(f_1(s))ds \approx \exp(f_1(t))\int_{0}^{t}\exp(f_1'(t)(s-t))ds
=\frac{\exp(f_1(t))}{f_1'(t)}(1-\exp(-tf_1'(t))).
\eeq
Using expression \eqref{y1exact}, we obtain for $t\in [0,t_1]$
\beq
y(t)\approx\left(\frac{1}{\tau_r}\frac{(1-\exp(-tf_1'(t)))}{f_1'(t)}\right)+C_1\exp(-f_1(t)), \label{y1approx}
\eeq
where
\beq
\beta_1=\cfrac{1}{\tau_r}+B_1LH_1 \text{ and }   C_1=\exp\left(-\cfrac{LH_1A_1}{\alpha_1}\right).
\eeq
We now compute the firing rate $h$: the first equation in system \eqref{simplifiedSyst} is
\beq
\tau\dot{h} = -h+Jxyh^+
\eeq
where the initial condition is $h(0) = \tilde{H}_1$.  We note that for numerical computations, the value of $\tilde{H}_1$ has to be much smaller than $H_1$ in order to guarantee that the facilitation and depression are immediately in the bursting state (Table \ref{approxParam}).\\
A direct integration leads to
\beq
h(t)=\tilde{H}_1\exp\left(-\frac{t}{\tau}+\frac{J}{\tau}\int_0^t x(s)y(s)ds\right). \label{h1int}
\eeq
We derive an explicit expression (appendix \ref{appendix1}) for the solution $h$ using \eqref{x1approx} and \eqref{y1approx} for $x$ and $y$ respectively.
\subsubsection*{Phase 2 $[t_1,t_2]$}
The second phase starts at $t_1$ where $H(t)=H_2$, the equations and the approximation are similar to the paragraph above. However we use the following initial conditions: $x(t_1^-)=x(t_1^+)$ and $y(t_1^-)=y(t_1^+)$. This yields for $t \in [t_1;t_2]$,
\beq
x(t)=A_2 e^{-\alpha_2 t}+B_2, \label{x2approx}
\eeq
where
\beq
\alpha_2=\cfrac{1}{\tau_f}+KH_2, \phantom{124553} A_2=(x(t_1^-)-B_2)e^{\alpha_2 t_1}, \phantom{124553} B_2=\cfrac{\cfrac{X}{\tau_f}+KH_2}{\alpha_2}.
\eeq
\beq \label{y2approx}
y(t)\approx\left(\frac{1}{\tau_r}\frac{(1-\exp(-(t-t_1)f_2'(t)))}{f_2'(t)}\right)+C_2\exp(-f_2(t)), 
\eeq
where
\beq
f_2(t)=\beta_2t-\cfrac{LA_2H_2}{\alpha_2}e^{-\alpha_2 t},
\eeq
\beq
\beta_2=\cfrac{1}{\tau_r}+B_2LH_2 \text{ and } C_2=y(t_1^-)\exp(f_2(t_1)).
\eeq
Finally, we use equation \eqref{T0values} for $T_0=T_{AHP}$ and $\tau_0=\tau_{mAHP}$ so that the voltage equation reduces to
\beq
\tau_{mAHP}\dot{h} = -(h-T_{AHP})+Jxy(h-T_{AHP})^+
\eeq
with the initial condition $h(t_1^+)=h(t_1^-)$. We obtain by a direct integration
\beq\label{h2solution}
h(t)=(h(t_1^-)-T_{AHP})\exp\left(-\frac{t-t_1}{\tau_{mAHP}}+\frac{J}{\tau_{mAHP}}\int_{t_1}^t x(s)y(s)ds\right)+T_{AHP},
\eeq
as detailed in appendix \ref{appendix2}.
\subsubsection*{Phase 3 $[t_2,\infty[$}
The recovery phase starts at time $t_2$ where $H(t)=0$. We use the following initial conditions: $x(t_2^-)=x(t_2^+)$ and $y(t_2^-)=y(t_2^+)$, leading for $t \geq t_2$ to the representation
\beq \label{x3approx}
x(t)=X+(x(t_2^-)-X)\exp\left(-\frac{t-t_2}{\tau_f}\right)
\eeq
\beq\label{y3approx}
y(t)=1+(y(t_2^-)-1)\exp\left(-\frac{t-t_2}{\tau_r}\right).
\eeq
Finally, when $t>t_2$, $h$ enters into a slow relaxation phase, (see relation \eqref{T0values}), where $T_0=0$ and $\tau_0=\tau_{sAHP}$, and the initial condition is $h(t_2^-)=h(t_2^+)$. A direct integration of equation \eqref{simplifiedSyst} leads to (see appendix \ref{appendix3} for the detailed solution)
\beq
h(t)=h(t_2^-)\exp\left(-\frac{t-t_2}{\tau_{sAHP}}+\frac{J}{\tau_{sAHP}}\int_{t_2}^t x(s)y(s)ds\right). \label{h3solution}
\eeq
\begin{figure}[h!] \centering
	\includegraphics[scale=0.8]{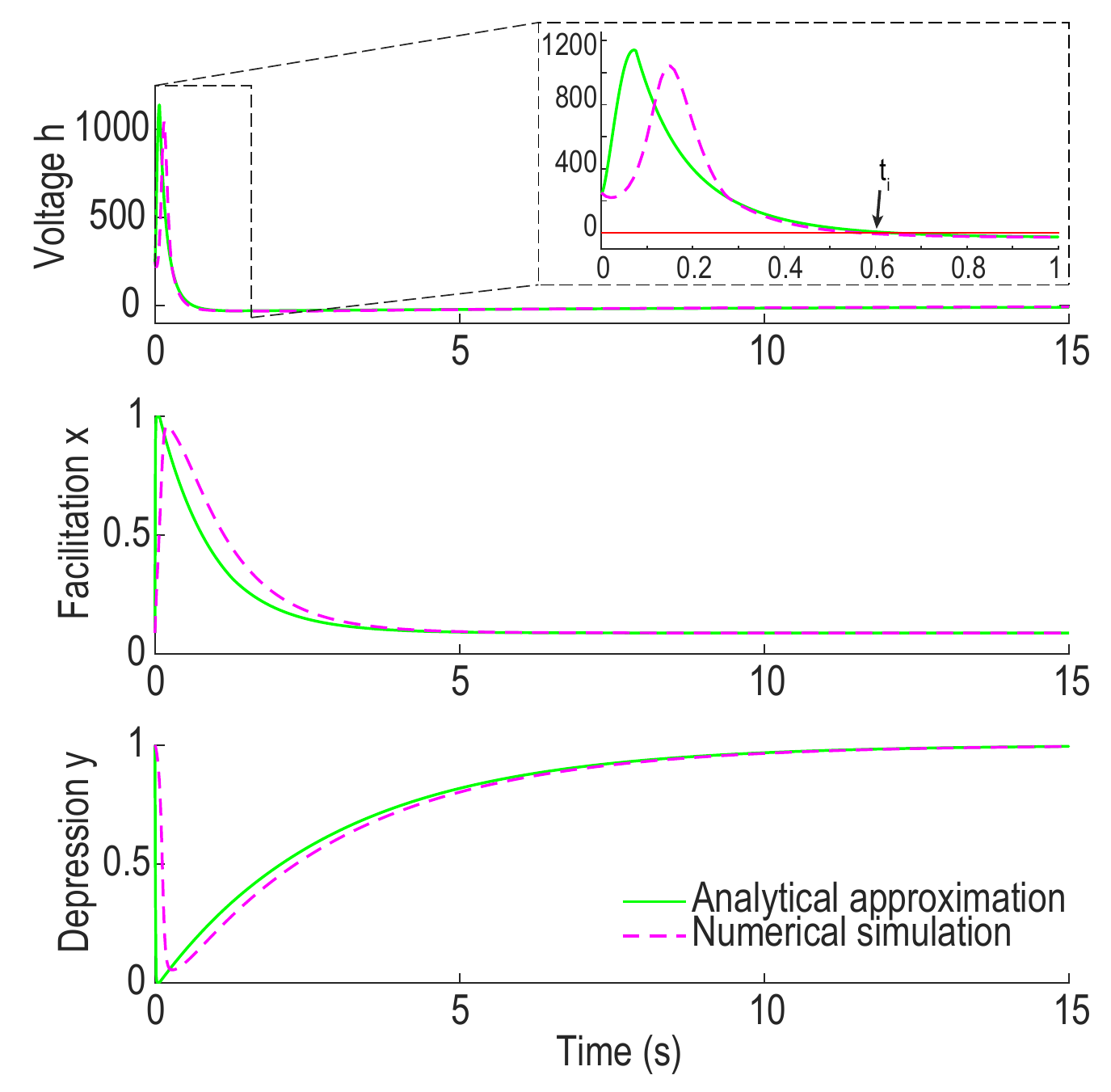}
	\caption{Analytical approximation (green) from formulas (\ref{h1int}, \ref{h2solution} and \ref{h3solution}) for the three phases for $h$ (upper), formulas (\ref{x1approx}, \ref{x2approx} and \ref{x3approx}) for $x$ (center) and (\ref{y1approx}, \ref{y2approx} and \ref{y3approx}) for $y$ (lower) vs exact solutions (dashed magenta) obtained numerically. With an inset of $h$ for $t \in [0,1]$, showing the burst duration $t_i$ such that $h(t_i)=0$ (red line: $h=0$).} \label{approx}
\end{figure}
\subsection{Identification of the termination times \boldmath $t_1 \text{ and } t_2$} \label{t1andt2}
\subsubsection*{End of phase 1}
Following burst activation, medium and slow $K^+$ channels start to be activated forcing the voltage to hyperpolarize. To account for the overall changes in the voltage dynamics due to this $K^+$ channels activation, we change the recovery timescale $\tau_0$ to $\tau_{mAHP}$ (equation \eqref{T0values}) and $H(t)$ to $H_2$ in \eqref{Hvalues} at time $t_1$. In practice the hyperpolarization initiation is defined in the region where $h$ is decreasing after reaching its maximum, as the first time $t_1$ such that $h(t_1)=h_0$  (expression \eqref{h1int}), leading to equation
\beq \label{t1exp}
\begin{split}
	\ds	t_1\cfrac{B_1J-\tau_r\beta_1}{\tau_r\beta_1J}
	-\cfrac{1}{\tau_rLH_1\alpha_1}\ln\left(\cfrac{1+\cfrac{LA_1H_1}{\beta_1}e^{-\alpha_1t_1}}{1+\cfrac{LA_1H_1}{\beta_1}}\right)+\cfrac{e^{-\alpha_1t_1}-1}{\alpha_1}\cfrac{LA_1H_1B_1}{\tau_r\beta_1^2}
	-\cfrac{e^{-2\alpha_1t_1}-1}{2\alpha_1}\cfrac{(LA_1H_1)^2B_1}{\tau_r\beta_1^3}+\\ \cfrac{e^{-(\beta_1+LA_1H_1)t_1}-1}{\beta_1+LA_1H_1}\cfrac{B_1(1-\tau_r\beta_1)}{\tau_r\beta_1}+
	\cfrac{e^{-(\alpha_1+\beta_1+LA_1H_1)t_1}-1}{\alpha_1+\beta_1+LA_1H_1}\cfrac{A_1(-\tau_r\beta_1^2+\beta_1-LH_1B_1)}{\tau_r\beta_1^2}\\
	+\cfrac{e^{-(2\alpha_1+\beta_1+LA_1H_1)t_1}-1}{2\alpha_1+\beta_1+LA_1H_1}\left(-\cfrac{LA_1^2H_1}{\tau_r\beta_1^2}+\cfrac{(LA_1H_1)^2B_1}{\tau_r \beta_1^3}\right)
	+\cfrac{e^{-(3\alpha_1+\beta_1+LA_1H_1)t_1}-1}{3\alpha_1+\beta_1+LA_1H_1}\cfrac{L^2A_1^3H_1^2}{\tau_r^2\beta_1^3}
	=\frac{\tau}{J}\ln\left(\frac{h_0}{H_1}\right).
\end{split}
\eeq
Equation \eqref{t1exp} is transcendental and cannot be solved explicitly. However, we will search for an approximated solution by
neglecting the exponential terms as shown by the range of our parameters (Tables \ref{tableParam} and \ref{approxParam}), leading to
\beq \label{t1}
\begin{split}	t_1=\frac{1}{\Gamma_1}\left(-\Gamma_2\ln\left(\cfrac{1}{1+\Gamma_3}\right)+\frac{\Gamma_4}{\alpha_1}+\frac{\Gamma_5}{\beta_1+LA_1H_1}+\frac{\Gamma_6}{\alpha_1+\beta_1+LA_1H_1}+\frac{\Gamma_7}{2\alpha_1+\beta_1+LA_1H_1}\right. \\ \left. +\frac{\Gamma_8}{3\alpha_1+\beta_1+LA_1H_1}+\frac{\Gamma_9}{2\alpha_1}+\frac{\tau}{J}\ln\left(\frac{h_0}{\tilde{H}_1}\right)\right),
\end{split}
\eeq
where
\beq \label{gammas}
\begin{array}{l}
	\Gamma_1 = \cfrac{B_1J-\tau_r\beta_1}{\tau_r\beta_1J}, \,\,
	\Gamma_2 = -\cfrac{1}{\tau_rLH_1\alpha_1}, \,\,
	\Gamma_3 = \cfrac{LA_1H_1}{\beta_1}, \,\,
	\Gamma_4 = \cfrac{LA_1H_1B_1}{\tau_r\beta_1^2},\,\,
	\Gamma_5 = \cfrac{B_1(1-\tau_r\beta_1e^{\frac{LA_1H_1}{\alpha_1}})}{\tau_r\beta_1},\\
	\Gamma_6 = \cfrac{A_1(-\tau_r\beta_1^2e^{\frac{LA_1H_1}{\alpha_1}}+\beta_1-LH_1B_1)}{\tau_r\beta_1^2},\,
	\Gamma_7 =\cfrac{LA_1^2H_1}{\tau_r\beta_1^2} \left(\cfrac{LH_1B_1}{\beta_1}-1\right),\,
	\Gamma_8 = \cfrac{L^2A_1^3H_1^2}{\tau_r^2\beta_1^3} \, \text{ and }
	\Gamma_9 = -\cfrac{(LA_1H_1)^2B_1}{\tau_r\beta_1^3}.
\end{array}
\eeq
In order to grasp the respective influence of the network parameters $J$, $K$ and $L$ on $t_1$, we rewrite formula \eqref{t1} by using the numerical values of all the other parameters (Table \ref{tableParam}), yielding (see appendix \ref{intParamSimple} relation \eqref{gammasSimple} for the intermediate formulas)
\beq \label{t1simple}
t_1(J) \approx \cfrac{29J}{2.9LH_1-J}\left(F(K,L)+\frac{\tau}{J}\ln\left(\frac{h_0}{\tilde{H}_1}\right)\right),
\eeq
where  $J\in [3-6]$ and
\beq
F(K,L)=\exp\left(-0.9\cfrac{L}{K}\right)\cfrac{0.01L+K}{0.1L+K}.
\eeq
With parameters of table \ref{tableParam}, $t_1 \approx 100$ ms, suggesting that the medium and slow $K^+$ channels start to be activated quite early following burst initiation.
\subsubsection*{End of phase 2} \label{t2}
The second phase is dominated by hyperpolarization and ends when the voltage reaches asymptotically its minimum. In practice we introduce a threshold $h_{AHP}$ so that when the condition $h(t_2)=h_{AHP}$ is satisfied (expression \eqref{h2solution}), we switch into the third phase (see \eqref{Hvalues} and \eqref{T0values}). This leads to equation
\beqq
\begin{split}
	(t_2-t_1)\cfrac{B_2J-\tau_r\beta_2}{\tau_r\beta_2J}
	-\cfrac{1}{\tau_rLH_2\alpha_2}\ln\left(\cfrac{1+\cfrac{LA_2H_2}{\beta_2}e^{-\alpha_2t_1}e^{-\alpha_2(t_2-t_1)}}{1+\cfrac{LA_2H_2}{\beta_2}e^{-\alpha_2t_1}}\right)+\cfrac{e^{-\alpha_2(t_2-t_1)}-1}{\alpha_2}e^{-\alpha_2t_1}\cfrac{LA_2H_2B_2}{\tau_r\beta_2^2}\\
	-\cfrac{e^{-2\alpha_2(t_2-t_1)}-1}{2\alpha_2}e^{-2\alpha_2t_1}\cfrac{(LA_2H_2)^2B_2}{\tau_r\beta_2^3}+ \cfrac{e^{-(\beta_2+LA_2H_2)(t_2-t_1)}-1}{\beta_2+LA_2H_2}e^{-(\beta_2+LA_2H_2)t_1}\cfrac{B_2(1-C_2e^{\frac{LA_2H_2}{\alpha_2}}\tau_r\beta_2)}{\tau_r\beta_2}\\
	\cfrac{e^{-(\alpha_2+\beta_2+LA_2H_2)(t_2-t_1)}-1}{\alpha_2+\beta_2+LA_2H_2}e^{-(\alpha_2+\beta_2+LA_2H_2)t_1}\cfrac{A_2(-C_2e^{\frac{LA_2H_2}{\alpha_2}}\tau_r\beta_2^2+\beta_2-LH_2B_2)}{\tau_r\beta_2^2}\\
	+\cfrac{e^{-(2\alpha_2+\beta_2+LA_2H_2)(t_2-t_1)}-1}{2\alpha_2+\beta_2+LA_2H_2}e^{-(2\alpha_2+\beta_2+LA_2H_2)t_1}\left(-\cfrac{LA_2^2H_2}{\tau_r\beta_2^2}+\cfrac{(LA_2H_2)^2B_2}{\tau_r \beta_2^3}\right)\\
	+\cfrac{e^{-(3\alpha_2+\beta_2+LA_2H_2)(t_2-t_1)}-1}{3\alpha_2+\beta_2+LA_2H_2}e^{-(3\alpha_2+\beta_2+LA_2H_2)t_1}\cfrac{L^2A_2^3H_2^2}{\tau_r^2\beta_2^3}
	=\frac{\tau_{mAHP}}{J}\ln\left(\cfrac{h_{AHP}-T_{AHP}}{h_0-T_{AHP}}\right).
\end{split}
\eeqq
Contrary to the equivalent expression \eqref{t1exp}, all terms are of the same order and thus we cannot neglect any of them. To estimate the value of $t_2$, we solve now numerically the transcendental equation
\beq \label{t2equation}
\begin{split}
	\Lambda_1(t_2-t_1) +\Lambda_2\ln\left(\frac{1+\Lambda_3e^{\ds-\alpha_2(t_2-t_1)}}{1+\Lambda_3}\right)+\Lambda_4\cfrac{e^{\ds-\alpha_2(t_2-t_1)}-1}{\alpha_2}+\Lambda_5\cfrac{e^{\ds-(\beta_2+LA_2H_2)(t_2-t_1)}-1}{\beta_2+LA_2H_2} \\
	+\Lambda_6\cfrac{e^{\ds-(\alpha_2+\beta_2+LA_2H_2)(t_2-t_1)}-1}{\alpha_2+\beta_2+LA_2H_2}
	+\Lambda_7\cfrac{e^{\ds-(2\alpha_2+\beta_2+LA_2H_2)(t_2-t_1)}-1}{2\alpha_2+\beta_2+LA_2H_2}\\
	+\Lambda_8\cfrac{e^{\ds-(3\alpha_2+\beta_2+LA_2H_2)(t_2-t_1)}-1}{3\alpha_2+\beta_2+LA_2H_2}+\Lambda_9\cfrac{e^{\ds-2\alpha_2(t_2-t_1)}-1}{2\alpha_2}-\frac{\tau_{mAHP}}{J}\ln\left(\cfrac{h_{AHP}-T_{AHP}}{h_0-T_{AHP}}\right)=0,
\end{split}
\eeq
where
\beq \label{lambdas}
\begin{array}{l}
	\Lambda_1 = \cfrac{B_2J-\tau_r\beta_2}{\tau_r\beta_2J}, \,
	\Lambda_2 = -\cfrac{1}{\tau_rLH_2\alpha_2}, \,
	\Lambda_3 = \cfrac{LA_2H_2}{\beta_2}e^{\ds-\alpha_2t_1}, \,
	\Lambda_4 = \cfrac{LA_2H_2B_2}{\tau_r\beta_2^2}e^{\ds-\alpha_2t_1}, \, \\
	\Lambda_5 = \cfrac{B_2(1-\tau_r\beta_2C_2e^{\frac{LA_2H_2}{\alpha_2}})}{\tau_r\beta_2}e^{\ds-(\beta_2+LA_2H_2)t_1},\\
	\Lambda_6 = \cfrac{A_2(-C_2e^{\frac{LA_2H_2}{\alpha_2}}\tau_r\beta_2^2+\beta_2-LH_2B_2)}{\tau_r\beta_2^2}e^{\ds-(\alpha_2+\beta_2+LA_2H_2)t_1}, \,\, \\
	\Lambda_7 = \cfrac{LA_2^2H_2}{\tau_r\beta_2^2}\left(\cfrac{LH_2B_2}{\beta_2}-1\right)e^{\ds-(2\alpha_2+\beta_2+LA_2H_2)t_1},\\
	\Lambda_8 = \cfrac{L^2A_2^3H_2^2}{\tau_r^2\beta_2^3}e^{\ds-(3\alpha_2+\beta_2+LA_2H_2)t_1} \,\text{ and }\,
	\Lambda_9 = -\cfrac{(LA_2H_2)^2B_2}{\tau_r\beta_2^3}e^{\ds-2\alpha_2t_1}.
\end{array}
\eeq
The time $t_2$ depends on $J$ and we obtain a numerical approximation for $J \in [2.95,5.25]$ by fitting a rational function of the same form as the one we obtained for $t_1$:
\beq\label{t2simple}
t_2(J) \approx \cfrac{43.96J+42.29}{-4.28J+130.4}.
\eeq
With our parameters we obtain $t_2 \approx 2$ s (appendix \ref{t2functionJ}). Finally, we compare the analytical approximation for $h$, $x$ and $y$ (dashed magenta) with the exact solution obtained using numerical simulations (green) in fig. \ref{approx}. In these computations, we have chosen $H_1 = 8000$ and $H_2 =-1$ such that the analytical approximations for $x$ and $y$ fit well their numerical solutions. Even though the burst peak is slightly before, with a shift of $\approx 100 $ms for the approximated system compared to the unperturbed solution (fig. \ref{approx} inset), the decreasing phase is similar for both. We use this numerical agreement to justify that we use the approximated system to compute the burst duration, by finding the time $t_i$ such as $h(t_i) = 0$ and in this time range the analytical approximation fits well the numerical solution.
\subsection{Bursting and AHP durations} \label{paramInfluenceOnBDAHP}
\subsubsection*{Bursting duration}
The burst duration is defined from the voltage jump at time $t=0$ to $h(t)=H_1$ and ends when $h(t_i)=0$ for the first time. In practice, we use expression \eqref{h2solution} as in section \ref{t1andt2} for the end of phase 2 however, here $t_i-t_1$ is small enough to allow us to use Taylor expansions to second order leading to the quadratic equation
\beq \label{tiequation}
\tilde{\Lambda}(t_i-t_1)^2+\Lambda(t_i-t_1)-\frac{\tau_{mAHP}}{J}\ln\left(\cfrac{-T_{AHP}}{h_0-T_{AHP}}\right)=0,
\eeq
where
\beqq
\begin{split}
	\tilde{\Lambda} =
	\left(\frac{\Lambda_2\Lambda_3\alpha_2^2}{2(1+\Lambda_3)^2}+\frac{1}{2}\big(\alpha_2\Lambda_4+(\beta_2+LA_2H_2)\Lambda_5+(\alpha_2+\beta_2+LA_2H_2)\Lambda_6+(2\alpha_2+\beta_2+LA_2H_2)\Lambda_7 \right.\\
	\left. +(3\alpha_2+\beta_2+LA_2H_2)\Lambda_8+2\alpha_2\Lambda_9\big) \vphantom{\frac{1}{2}}\right),
\end{split}
\eeqq
and
\beqq
\Lambda = \left(-\frac{\Lambda_2\Lambda_3\alpha_2}{1+\Lambda_3}+\Lambda_1-\Lambda_4-\Lambda_5-\Lambda_6-\Lambda_7-\Lambda_8-\Lambda_9\right)
\eeqq
We keep the positive root
\beq \label{BurstDuration}
t_i \approx t_1(J)+\cfrac{-\Lambda(J,K,L,H_2,t_1(J)) -\sqrt{\Lambda^2(J,K,L,H_2,t_1(J)) +4\tilde{\Lambda}(K,L,H_2,t_1(J)) \cfrac{\tau_{mAHP}}{J}\ln\left(\cfrac{-T_{AHP}}{h_0-T_{AHP}}\right)}}{2\tilde{\Lambda}}.
\eeq
Similarly as for $t_1$, we give simplified formulas for $\Lambda$ and $\tilde{\Lambda}$ depending only on the parameters $J$, $K$, $L$ and the threshold $H_2$ such as
\beq\label{LamdaTildSimple}
\tilde{\Lambda}(K,L,H_2,t_1(J))=\cfrac{-0.08-0.49KH_2-0.66LH_2+0.06t_1(J)+0.54KH_2t_1(J)+0.1LH_2t_1(J)}{0.27+2.26KH_2+1.9L H_2},
\eeq
and
\beq
\Lambda(J,K,L,H_2,t_1(J)) = \cfrac{A(K,L,H_2)+\left(B(K,L,H_2)+C(K,L,H_2)t_1(J)\right)J}{J(\tilde{A}+\tilde{B}J)},
\eeq
where
\beq
\begin{array}{r c l}
A(K,L,H_2) & = & 33.3-246KH_2-143LH_2\\
B(K,L,H_2) & = & 27.2-122KH_2-106LH_2\\
C(K,L,H_2) & = & 11.6+86.3KH_2+49.8LH_2\\
\tilde{A} & = & 33.3\\
\tilde{B}(K,L,H_2) & = & 246KH_2+143LH_2
\end{array}
\eeq
The simplified formulas for the intermediary parameters $\Lambda_1 - \Lambda_9$ are given in appendix \ref{intParamSimple} formula \eqref{lambdasSimple}.\\
Using parameters from Table \ref{tableParam} and Table \ref{approxParam}, we obtain $t_i \approx 0.6$ s, which is comparable to the bursting times observed in experimental data \cite{Rouach_CxKO}, and from our numerical simulations in the noiseless case (fig. \ref{PhaseSpace}D).
\subsubsection*{AHP duration}
The AHP starts at time $t_i$ computed above, however using expression \eqref{h3solution} the termination time to reach $h(t_e) = 0$ would be infinite. Thus, we introduce a threshold $\epsilon$ and define the end of AHP $t_e$ such as $h(t_e)=\epsilon$. In practice, the value $\epsilon$ can be estimated from the amplitude of the voltage fluctuations at equilibrium. We obtain from expression \eqref{h3solution}
\beqq
\begin{split}
	\left(-\frac{1}{J}+X\right)(t_e-t_2)-\tau_rX(y(t_2)-1)\left(e^{\ds-\frac{t_e-t_2}{\tau_r}}-1\right)-\tau_f(x(t_2)-X)\left(e^{\ds-\frac{t_e-t_2}{\tau_f}}-1\right)\\
	-\cfrac{(y(t_2)-1)(x(t_2)-X)\tau_f\tau_r}{\tau_f+\tau_r}\left(e^{\ds-(t-t_2)\frac{\tau_f+\tau_r}{\tau_f\tau_r}}-1\right)=\frac{\tau_{sAHP}}{J}\ln\left(\frac{\epsilon}{h(t_2)}\right)
\end{split}
\eeqq
because $t_e-t_2$ is large enough, we neglect the exponential terms so that
\beqq
(t_e-t_2)\left(X-\frac{1}{J}\right)+\tau_rX(y(t_2)-1)+\tau_f(x(t_2)-X)+\cfrac{(x(t_2)-X)(y(t_2)-1)\tau_f\tau_r}{\tau_f+\tau_r}=\frac{\tau_{sAHP}}{J}\ln\left(\frac{\epsilon}{h(t_2)}\right),
\eeqq
leading to
\beqq
t_e = t_2+\left(\frac{\tau_{sAHP}}{J}\ln\left(\frac{\epsilon }{h(t_2)}\right)-\tau_rX(y(t_2)-1)-\tau_f(x(t_2)-X)-\cfrac{(x(t_2)-X)(y(t_2)-1)\tau_f\tau_r}{\tau_f+\tau_r}\right)\cfrac{J}{JX-1},
\eeqq
This simplifies to
\beq \label{AHPduration}
t_e \approx t_2(J)+\left(\frac{\tau_{sAHP}}{J}\ln\left(\frac{h(t_2) }{\epsilon}\right)+0.26\right)\cfrac{J}{1-JX},
\eeq
using the approximated value of $t_2$ we obtain
\beqq
t_e \approx \cfrac{43.96J+42.29}{-4.28J+130.4}+\left(\frac{\tau_{sAHP}}{J}\ln\left(\frac{ h(t_2)}{\epsilon}\right)+0.26\right)\cfrac{J}{1-JX}.
\eeqq
Using the parameter values from Table \ref{tableParam} and Table \ref{approxParam} we obtain $t_e \approx 15.4$ s and $\Delta_{AHP} = t_e-t_i \approx 14.3$ s, which is coherent with the durations obtained from the numerical simulations (fig. \ref{FigureModel}D), as well as classical AHP durations found in the literature \cite{AHP_review}.
\subsubsection{Study of parameter influence on burst and AHP durations} \label{paramInfAnalytical}
To evaluate the influence of the main parameters on the bursting and AHP durations we plotted these times vs the recovery timescales $\tau_{mAHP}$ and $\tau_{sAHP}$, the hyperpolarization level $T_{AHP}$ and the arbitrary thresholds $h_0$, $\tilde{H}_1$, $h_{AHP}$ and $\epsilon$. First, the burst duration that varies between 0.5 and 3s, is an increasing function of $\tau_{mAHP}$ and does not depend much on $T_{AHP}$ in the range $[-15;-40]$ (fig. \ref{paramInfluence}A). In addition, the AHP duration increases with $\tau_{sAHP}$, but in a larger range from 9 to 35s. However, the hyperpolarization level $T_{AHP}$ has a larger influence on this duration (fig. \ref{paramInfluence}B). To verify that the arbitrary thresholds that we use do not influence much the burst and AHP durations, we plotted them in fig. \ref{paramInfluence}C-F with respect to the phase 1 termination threshold $h_0$, the phase 2 termination threshold $h_{AHP}$, the duration of phase 1 $t_1$ and the AHP termination threshold $\epsilon$, respectively. These figures show that there is almost no dependency with respect to $\tilde{H}_1$ and $T_{AHP}$, as well as $h_0$ and $h_{AHP}$ due to the effect of the logarithmic term.
\begin{figure} \center
\includegraphics[scale=0.8]{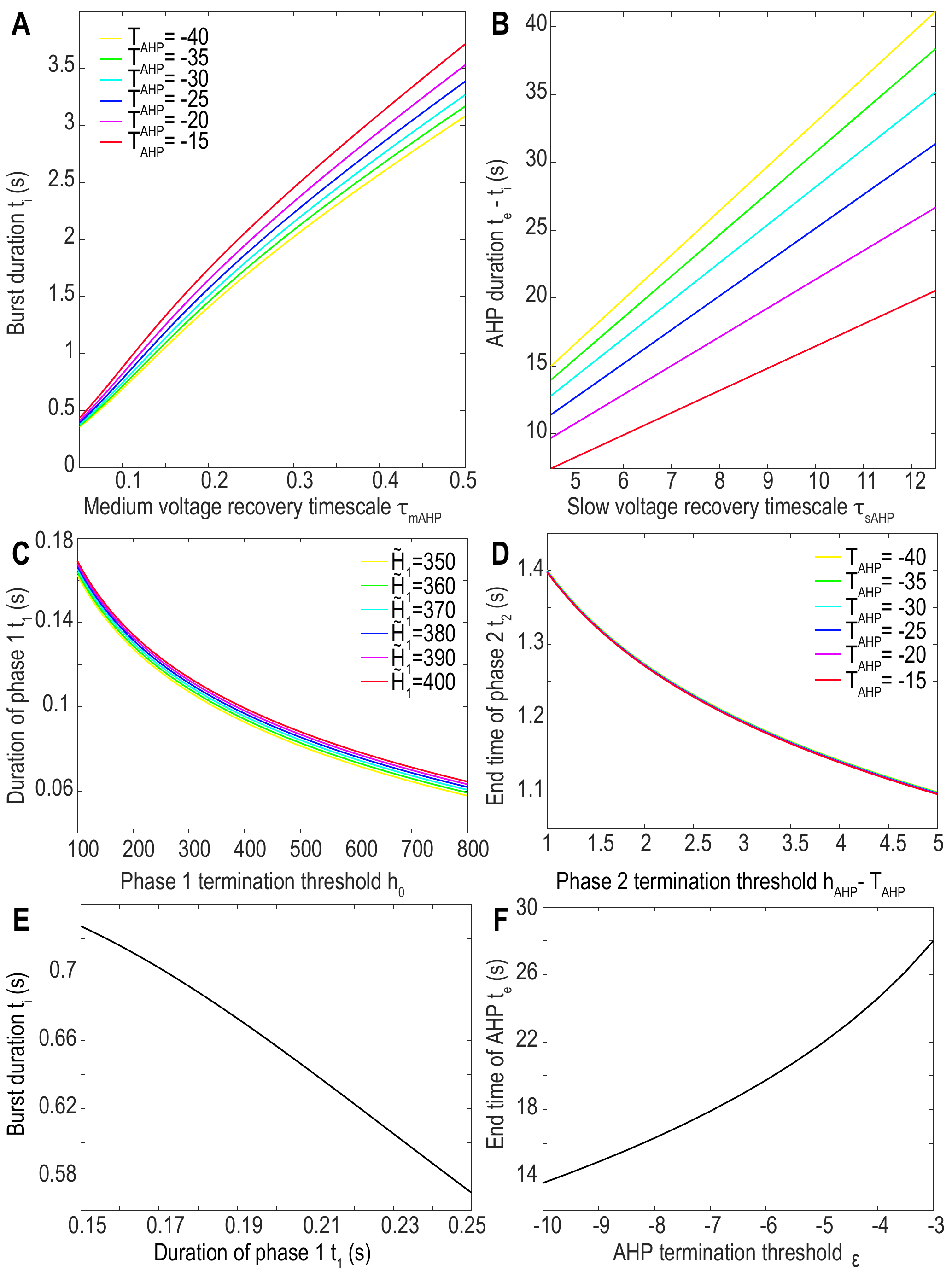}
\caption{{\bf Parameter influence on burst and AHP durations.} \textbf{A}. Evolution of the burst duration $t_i$ as a function of the medium timescale $\tau_{mAHP}$ for multiple values of the hyperpolarization level $T_{AHP}$. \textbf{B}. Evolution of the AHP duration $t_e - t_i$ as a function of the slow timescale $\tau_{sAHP}$ for multiple values of the hyperpolarization level $T_{AHP}$. \textbf{C}. Duration of phase 1 $t_1$ as a function of its termination threshold $h_0$ for multiple values of the initial voltage value $h(0)=\tilde{H}_1$. \textbf{D}. End time of phase 2 $t_2$ as a function of its termination threshold $h_{AHP}$ (relatively to $T_{AHP}$) for multiple values of the hyperpolarization level $T_{AHP}$. \textbf{E}. Bursting duration as a function of $t_1$ for $\tau_{mAHP}=0.1s$. \textbf{F}. AHP duration as a function of the threshold $\epsilon$ for $\tau_{sAHP}=7.5s$ and $T_{AHP}=-30$.}\label{paramInfluence}
\end{figure}
\subsection{Burst and IBI durations vs J, K, L parameters} \label{infJKL}
\begin{figure} \centering
	\includegraphics[scale=0.8]{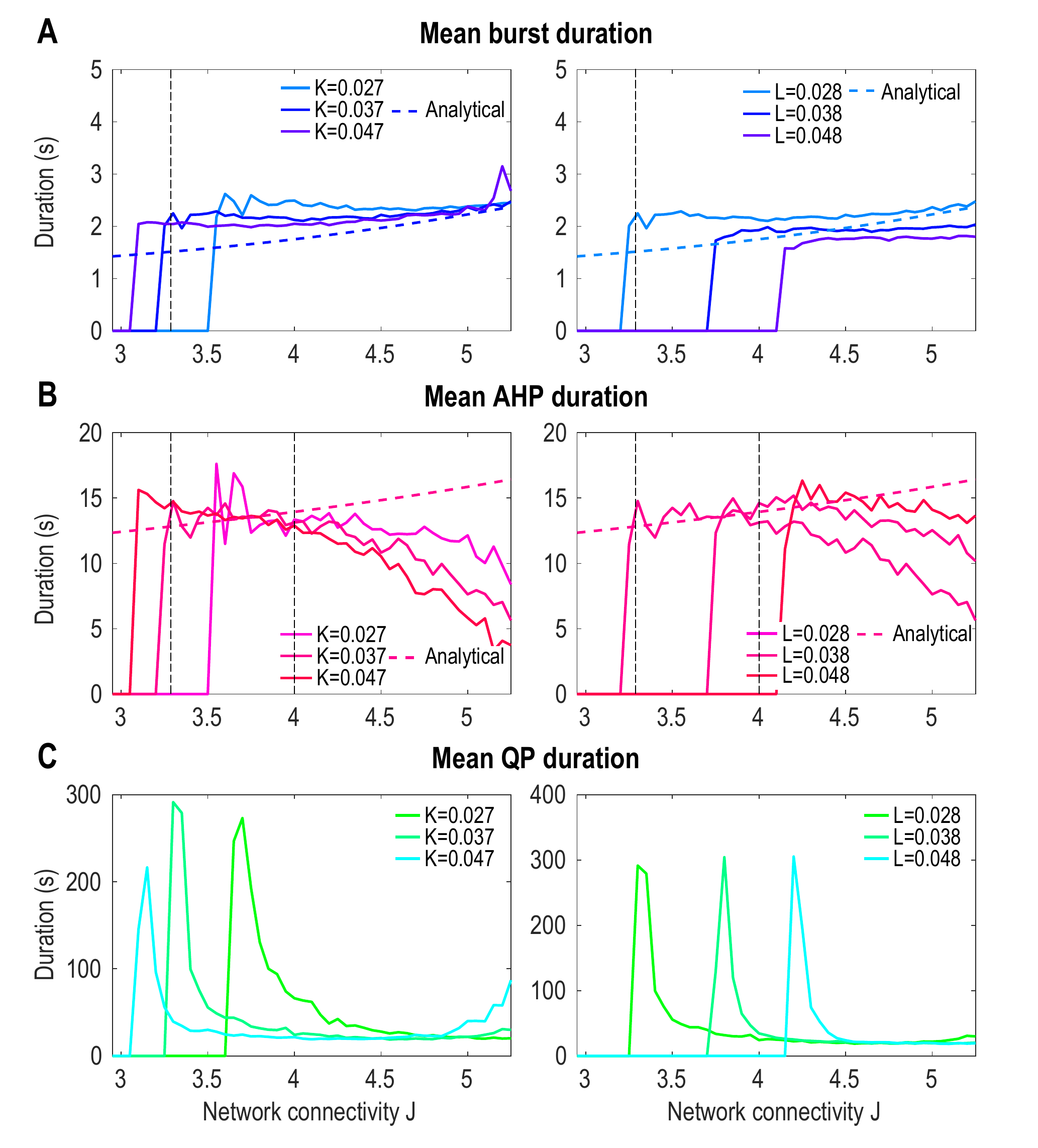}
	\caption{\textbf{Influence of the network connectivity J on bursting dynamics.} \textbf{A}. (resp.\textbf{B}, \textbf{C}) Mean burst (resp. AHP, QP) duration in seconds from 5000s simulations for $J$ varying from 2.95 to 5.25 and three values of $K$ (left) and $L$ (right) with a fixed noise level ($\sigma=6$) compared to the analytical result (dashed) obtained for $t_1$ \eqref{BurstDuration} (resp. $\Delta_{AHP}=t_e-t_i$ \eqref{AHPduration}) for $(K,L) = (0.037,0.028)$. The vertical black dashed lines show the range of validity for $J$.} \label{JKL}
\end{figure}
To study the influence of the network connectivity $J$ on burst, AHP and QP durations, we ran numerical simulations of the stochastic system \eqref{AHP_model}, where we varied $J$, as well as the facilitation and depression parameters $K$ and $L$. To determine the time distributions of burst and IBI, we segmented the traces obtained for 5000 seconds simulations with a noise amplitude $\sigma=6$ and computed the mean value of the bursts (fig. \ref{JKL}A), AHP (fig. \ref{JKL}B) and QP durations (fig. \ref{JKL}C). Interestingly, we observe two different regimes depending on the values of the parameters: no bursts ($J<3.05$ for $K= 0.047, L=0.028$; $J<3.2$ for $K= 0.037, L=0.028$; $J<3.5$ for $K= 0.027, L=0.028$; fig. \ref{JKL} left column, or $J<3.7$ for $K=0.037, L=0.038$ and $J<4.1$ for $K=0.037, L=0.048$, right column) and bursts followed by AHP (for higher values of $J$).\\
Interestingly, in the bursting regime, changing $J$ does not influence much the mean burst duration. However, AHP durations decreases as $J$ increases. Finally, QP durations reach a peak at the transition value of $J$ between the two regimes and then quickly decrease around QP $\approx 25s$. The mean burst durations obtained here are longer than the ones observed in fig. \ref{FigureModel}D, because in these simulations, we used $\sigma=6$ (vs $\sigma=3$ for fig. \ref{FigureModel}D). Indeed, the mean burst duration increases with the noise because, at the beginning of the burst, the deterministic part of the trajectory is still perturbed by the noise component, leading to a longer trajectory when the noise level is higher.\\
We also compared these mean durations to the ones obtained with the analytical formulas \eqref{BurstDuration} and \eqref{AHPduration} (fig. \ref{JKL}A, B dashed lines) with $K=0.037$ and $L=0.028$. To account for the difference in burst durations induced by the noise, we added a constant $c=0.8s$ to the burst durations obtained from our analytical formula.
The burst duration increases slowly with the network connectivity $J$, which is comparable to the numerical observations (for $J \geq 3.25$, black dashed line). We also compared the analytical AHP duration to the one observed with the numerical results: we obtain a good fit for a small range of $J$ (between $J \in [3.25,4]$, black dashed lines) but then the AHP value keeps increasing with our analytical result, whereas it is decreasing in the numerical simulations. This difference might be due to the effect of noise that modifies the deterministic behavior of the system.\\
To conclude, a sufficient connectivity level is necessary to generate bursting, however inside this regime, increasing the level of neuronal connectivity does not change much the bursting times.
\section*{Conclusion and discussion}
We present here a novel mean-field model of synaptic short-term plasticity for the voltage, depression and facilitation variables that now accounts for long AHP periods. This model generalizes the facilitation-depression model introduced in \cite{Tsodyks1997} and developed in \cite{Holcman_Tsodyks2006,Mongillo2008,DaoDuc2015,daoduc2014frontiers}. The AHP significantly increases the interburst duration by introducing a recovery phase after network bursting. When a Gaussian noise of small amplitude is added to the dynamics, it exhibits spontaneous bursts followed by AHP periods. We have studied here the distribution of bursts and of interbursts, decomposed in AHP and QP durations. Interestingly, we found that the distribution of bursts durations is quite concentrated (subsection \ref{ModelDescription}). To explain this property, we studied the three-dimensional phase-space of the dynamical system \eqref{AHP_model}, that contains one attractor and two saddle points. By computing numerically the two-dimensional stable manifold at one of the saddles, we found the distribution of exit points (on this manifold) when the initial point of the stochastic dynamics is located at the attractor. To compute this distribution we used two methods: 1) stochastic simulations, and 2) the method of characteristics to solve the FPE \eqref{fpe} in the limit of small noise. In both cases, we found a peaked distribution of exit points close to the saddle point, as predicted for two-dimensional stochastic systems \cite{SchussSIAM2002,OPT,Spivak1998}, summarized by expression \eqref{fluxFinal}. After the stochastic trajectories have crossed the separatrix, they follow an almost deterministic behavior, confirming that the distribution of exit points on the separatrix defines the spread of the distribution of burst durations.\\
We also derived here analytical formulas (subsection \ref{paramInfluenceOnBDAHP}) that reveal the influence of the parameters on burst and AHP durations. These computations can be used to calibrate the AHP parameters with respect to the expected values of burst and AHP durations, that could be measured experimentally. This model could thus be used to decipher the main mechanisms leading to changes in bursting and interburst dynamics, for example when the neuronal network is disrupted, during epilepsy or in the case of a glial network alteration \cite{Rouach_CxKO}.\\
Classical bursting models describe accurately the burst phase \cite{Izhikevitch2007,CoombesBressloff,Desroches2019,Ermentrout1986}, but interburst is often considered as the continuation in the phase-space of the deterministic trajectories. Here the interburst phase is composed of a deterministic refractory period, the AHP, followed by the escape from an attractor due to noise (subsection \ref{ModelDescription}). During successive bursts, trajectories are not reset at the attractor, but explore the region $B_-$ of non bursting trajectories. This exploration depends on the previous bursting trajectory. Thus, we expect a correlation between successive burst and interburst durations. This correlation may also depend on the amplitude of the voltage fluctuations. Finally, we predict that modifying the AHP duration could affect bursting, because it corresponds to a change in the attractor's position and dominates the effect of synaptic depression.
\newpage
\appendix
\section{Appendix: detailed computations of burst and AHP durations}
\subsection{Integral term of $h$ in phase 1}\label{appendix1}
To compute the integral in expression \eqref{h1int}, we split it into two parts:
\beqq
\ds\int_0^{t} x(s)y(s)ds = \underbrace{C_1\int_0^{t} (A_1e^{-\alpha_1s}+B_1)e^{-f_1(s)}}_I ds + \underbrace{\int_{0}^{t}(A_1e^{-\alpha_1s}+B_1)\left(\cfrac{1-e^{-sf_1'(s)}}{\tau_r f_1'(s)}\right)ds}_{II}.
\eeqq
We start by I:
\beqq
\ds I = \underbrace{C_1A_1\int_0^{t} e^{\ds-(\alpha_1+\beta_1)s+\frac{LA_1H_1}{\alpha_1}e^{-\alpha_1s}}ds}_{I_A} + \underbrace{C_1B_1\int_0^{t} e^{\ds-\beta_1s+\frac{LA_1H_1}{\alpha_1}e^{-\alpha_1s}}ds}_{I_B}.
\eeqq
Using a Taylor expansion at first order, $e^{-\alpha_1s}\approx 1-\alpha_1s$, we obtain
\beqq
I_A(t) \approx A_1C_1\int_0^{t}e^{\ds \frac{LA_1H_1}{\alpha_1}-\big(\alpha_1+\beta_1+LA_1H1\big)s}ds \approx -\cfrac{A_1\left(e^{\ds-(\alpha_1+\beta_1+LA_1H1)t}-1\right)}{\alpha_1+\beta_1+LA_1H1}
\eeqq
and
\beqq
I_B(t) \approx -\cfrac{B_1\left(e^{\ds-(\beta_1+LA_1H1)t}-1\right)}{\beta_1+LA_1H1}.
\eeqq
Similarly, we write $II = II_A+II_B$, where
\beqq
\begin{split}
II_A(t) = \cfrac{A_1}{\tau_r} \int_0^{t} \cfrac{e^{\ds-\alpha_1 s}\left(1 - e^{\ds-\beta_1 s-LA_1H_1se^{-\alpha_1 s}}\right)}{\beta_1 + LA_1H_1e^{\ds-\alpha_1 s}}ds \\
	\approx \cfrac{A_1}{\tau_r\beta_1} \left(\underbrace{\int_0^{t} \cfrac{e^{\ds-\alpha_1 s}}{1+\cfrac{LA_1H_1}{\beta_1}e^{\ds-\alpha_1 s}}ds}_{(i)} - \underbrace{\int_0^{t}\cfrac{e^{\ds-(\alpha_1+\beta_1+LA_1H_1)s}}{1 +\cfrac{LA_1H_1}{\beta_1}e^{\ds-\alpha_1 s}}ds}_{(ii)}\right).
\end{split}
\eeqq
For (i), using the change of variable $u=e^{-\alpha_1 s}$, we obtain
\beqq
(i)=-\cfrac{1}{\alpha_1}\int_{1}^{e^{-\alpha_1t}}\cfrac{du}{1+\cfrac{LA_1H_1}{\beta_11}u}=-\cfrac{\beta_1}{\alpha_1LA_1H_1}\ln\left(\cfrac{1+\cfrac{LA_1H_1}{\beta_1}e^{-\alpha_1t}}{1+\cfrac{LA_1H_1}{\beta_1}}\right)
\eeqq
For small $s$, $se^{\-\alpha_1s} \approx s$ and using the condition $\left|\cfrac{LA_1H_1}{\beta_1}\right| < 1$, we expand the denominator to second order to obtain
\beqq
\begin{split} (ii)\approx\int_{0}^{t}e^{\ds-(\alpha_1+\beta_1+LA_1H_1)s}\left(1-\cfrac{LA_1H_1}{\beta_1}e^{\ds-\alpha_1s}+\left(\cfrac{LA_1H_1}{\beta_1}\right)^2e^{\ds-2\alpha_1s}\right)ds\\
	\approx -\cfrac{e^{\ds-(\alpha_1+\beta_1+LA_1H_1)t}-1}{\alpha_1+\beta_1+LA_1H_1}+\cfrac{LA_1H_1}{\beta_1}\cfrac{e^{\ds-(2\alpha_1+\beta_1+LA_1H_1)t}-1}{2\alpha_1+\beta_1+LA_1H_1}\\
	-\left(\cfrac{LA_1H_1}{\beta_1}\right)^2\cfrac{e^{\ds-(3\alpha_1+\beta_1+LA_1H_1)t}-1}{3\alpha_1+\beta_1+LA_1H_1}.
\end{split}
\eeqq
Finally,
\beqq
\begin{split}
	II_A(t)\approx -\cfrac{1}{\tau_rLH_1\alpha_1}\ln\left(\cfrac{1+\cfrac{LA_1H_1}{\beta_1}e^{\ds-\alpha_1t}}{1+\cfrac{LA_1H_1}{\beta_1}}\right)+\cfrac{A_1}{\tau_r\beta_1}\cfrac{e^{\ds-(\alpha_1+\beta_1+LA_1H_1)t}-1}{\alpha_1+\beta_1+LA_1H_1}\\
	-\cfrac{LA_1^2H_1}{\tau_r\beta_1^2}\cfrac{e^{\ds-(2\alpha_1+\beta_1+LA_1H_1)t}-1}{2\alpha_1+\beta_1+LA_1H_1} + \cfrac{L^2A_1^3H_1^2}{\tau_r^2\beta_1^3}\cfrac{e^{\ds-(3\alpha_1+\beta_1+LA_1H_1)t}-1}{3\alpha_1+\beta_1+LA_1H_1}.
\end{split}
\eeqq
Similarly, we obtain the following expression for
\beqq
\begin{split}
II_B(t)\approx \cfrac{B_1}{\tau_r\beta_1}\int_0^t \left(1-\cfrac{LA_1H_1}{\beta_1}e^{\ds-\alpha_1s}+\left(\cfrac{LA_1H_1}{\beta_1}\right)^2e^{\ds-2\alpha_1s}-e^{\ds-(\beta_1+LA_1H_1)s} \right.\\
\left.+\cfrac{LA_1H_1}{\beta_1}e^{\ds-(\alpha_1+\beta_1+LA_1H_1)s}-\left(\cfrac{LA_1H_1}{\beta_1}\right)^2e^{\ds-(2\alpha_1+\beta_1+LA_1H_1)s}\right) ds \end{split}
\eeqq
\beqq
\begin{split}
	II_B\approx \cfrac{B_1}{\tau_r\beta_1}\left(t+\cfrac{LA_1H_1}{\beta_1}\cfrac{e^{\ds-\alpha_1t}-1}{\alpha_1}-\left(\cfrac{LA_1H_1}{\beta_1}\right)^2\cfrac{e^{\ds-2\alpha_1t}-1}{2\alpha_1}+\cfrac{e^{\ds-(\beta_1+LA_1H_1)t}-1}{\beta_1+LA_1H_1}\right.\\
	\left. -\cfrac{LA_1H_1}{\beta_1}\cfrac{e^{\ds-(\alpha_1+\beta_1+LA_1H_1)t}-1}{\alpha_1+\beta_1+LA_1H_1}+\left(\cfrac{LA_1H_1}{\beta_1}\right)^2\cfrac{e^{\ds-(2\alpha_1+\beta_1+LA_1H_1)t}-1}{2\alpha_1+\beta_1+LA_1H_1}\right).
\end{split}
\eeqq
\subsection{Integral term of $h$ in phase 2}\label{appendix2}
Our goal is now to compute expression \eqref{h2solution}. We decompose it into four parts:
\beqq
\int_{t_1}^t x(s)y(s)ds = I_A+I_B+II_A+II_B.
\eeqq
All computations and approximations are similar except that we integrate between $t_1$ and $t$. We obtain
\beqq
I_A(t) \approx -\cfrac{A_2C_2e^{\frac{LA_2H_2}{\alpha_2}}\left(e^{\ds-(\alpha_2+\beta_2+LA_2H_2)t}-e^{\ds-(\alpha_2+\beta_2+LA_2H_2)t_1}\right)}{\alpha_2+\beta_2+LA_2H_2}
\eeqq
\beqq
I_B(t) \approx -\cfrac{B_2C_2e^{\ds\frac{LA_2H_2}{\alpha_2}}\left(e^{\ds-(\beta_2+LA_2H2)t}-e^{\ds-(\beta_2+LA_2H_2)t_1}\right)}{\beta_2+LA_2H_2}
\eeqq
\beqq
\begin{split}
	II_A(t)\approx -\cfrac{1}{\tau_rLH_2\alpha_2}\ln\left(\cfrac{1+\cfrac{LA_2H_2}{\beta_2}e^{\ds-\alpha_2t}}{1+\cfrac{LA_2H_2}{\beta_2}e^{\ds-\alpha_2t_1}}\right)\\
	+\cfrac{A_2}{\tau_r\beta_2}\cfrac{e^{\ds-(\alpha_2+\beta_2+LA_2H_2)t}-e^{\ds-(\alpha_2+\beta_2+LA_2H_2)t_1}}{\alpha_2+\beta_2+LA_2H_2}\\
	-\cfrac{LA_2^2H_2}{\tau_r\beta_2^2}\cfrac{e^{\ds-(2\alpha_2+\beta_2+LA_2H_2)t}-e^{\ds-(2\alpha_2+\beta_2+LA_2H_2)t_1}}{2\alpha_2+\beta_2+LA_1H_2}\\
	+\cfrac{L^2A_2^3H_2^2}{\tau_r^2\beta_2^3}\cfrac{e^{\ds-(3\alpha_2+\beta_2+LA_2H_2)t}-e^{\ds-(3\alpha_2+\beta_2+LA_2H_2)t_1}}{3\alpha_2+\beta_2+LA_1H_2}
\end{split}
\eeqq
\beqq
\begin{split}
	II_B(t)\approx \cfrac{B_2}{\tau_r\beta_2}\left(t-t_1+\cfrac{LA_2H_2}{\beta_2}\cfrac{e^{\ds-\alpha_2t}-e^{\ds-\alpha_2t_1}}{\alpha_2}-\left(\cfrac{LA_2H_2}{\beta_2}\right)^2\cfrac{e^{\ds-2\alpha_2t}-e^{\ds-2\alpha_2t_1}}{2\alpha_2} \right.\\ \left.
	+\cfrac{e^{\ds-(\beta_2+LA_2H_2)t}-e^{\ds-(\beta_2+LA_2H_2)t_1}}{\beta_2+LA_2H_2}
	-\cfrac{LA_2H_2}{\beta_2}\cfrac{e^{\ds-(\alpha_2+\beta_2+LA_2H_2)t}-e^{\ds-(\alpha_2+\beta_2+LA_2H_2)t_1}}{\alpha_2+\beta_2+LA_2H_2} \right.\\ \left.
	+\left(\cfrac{LA_2H_2}{\beta_2}\right)^2\cfrac{e^{\ds-(2\alpha_2+\beta_2+LA_2H_2)t}-e^{\ds-(2\alpha_2+\beta_2+LA_2H_2)t_1}}{2\alpha_2+\beta_2+LA_2H_2}\right).
\end{split}
\eeqq
\subsection{Integral term of $h$ in phase 3}\label{appendix3}
Similarly as in phases 1 and 2 we compute the integral in expression \eqref{h3solution} and obtain
\beqq
\begin{split}
	\int_{t_2}^{t}x(s)y(s)ds = \int_{t_2}^{t}(X+X(y(t2^-)-1)e^{\ds\frac{t_2-s}{\tau_r}}+(x(t2^-)-X)e^{\ds\frac{t_2-s}{\tau_f}}\\
	+(y(t2^-)-1)(x(t2^-)-X)e^{\ds(t_2-s)(\frac{1}{\tau_f}+\frac{1}{\tau_r})}ds
\end{split}
\eeqq
\beqq
\begin{split}	=X(t-t_2)-\tau_rX(y(t2^-)-1)(e^{\ds-\frac{t-t_2}{\tau_r}}-1)-\tau_f(x(t2^-)-X)\left(e^{\ds-\frac{t-t_2}{\tau_f}}-1\right) \\ -\cfrac{(y(t2^-)-1)(x(t2^-)-X)\tau_f\tau_r}{\tau_f+\tau_r}\left(e^{\ds-(t-t_2)\frac{\tau_f+\tau_r}{\tau_f\tau_f}}-1\right).
\end{split}
\eeqq

\newpage
\subsection{Numerical values of intermediate and approximation parameters}
~\\~\\
\begin{center}
	\begin{tabular}{l l l l}
		& Parameters & Values \\
		\hline
		$\Gamma_1$ & -0.24 & $\Lambda_1$ & -0.18\\
		$\Gamma_2$ & 5.2.10$^{-6}$& $\Lambda_2$ & 11.47\\
		$\Gamma_3$ & -0.91& $\Lambda_3$ & -0.077\\
		$\Gamma_4$ & 1.4.10$^{-3}$& $\Lambda_4$ & 4.4.10$^{-3}$\\
		$\Gamma_5$ & -0.50& $\Lambda_5$ & 0.054\\
		$\Gamma_6$ & 0.46& $\Lambda_6$ & 0.89\\
		$\Gamma_7$ & 1.9.10$^{-6}$& $\Lambda_7$ & 0.07\\
		$\Gamma_8$ & 4.0.10$^{-4}$& $\Lambda_8$ & 1.8.10$^{-3}$\\
		$\Gamma_9$ & 1.3.10$^{-3}$& $\Lambda_9$ & 2.8.10$^{-3}$\\
		\hline 		
	\end{tabular}
\end{center}\captionof{table}{Intermediate parameters}\label{interParam}
~\\~\\
\begin{center}
	\begin{tabular}{l l l}
		& Parameters & Values \\
		\hline
		$H_1$ & Approximation of $h$ for $x$ and $y$ during phase 1 &8000\\
		$\tilde{H}_1$ & Initial value of $h$ & 250\\
		$H_2$ & Approximation of $h$ for $x$ and $y$ during phase 2 & -1\\
		$h_0$ & End of phase 1 threshold & 400\\
		$h_{AHP}$& End of phase 2 threshold & -29\\
		$\epsilon$ & End of AHP threshold & -5\\
		$t_1$ & End of phase 1 time & 200ms\\
		$t_2$ & End of phase 2 time & 1.37s\\
		$A_1$ & Approximation of $x$ on phase 1 parameter & -0.91\\
		$B_1$ & Approximation of $x$ on phase 1 parameter & 0.99\\
		$C_1$ & Approximation of $y$ on phase 1 parameter & 1.98\\
		$\alpha_1$ & Approximation of $x$ on phase 1 parameter & 297Hz\\
		$\beta_1$ & Approximation of $y$ on phase 1 parameter & 224Hz\\
		$A_2$ & Approximation of $x$ on phase 2 parameter & 1.16\\
		$B_2$ &  Approximation of $x$ on phase 2 parameter & 0.06\\
		$C_2$ & Approximation of $y$ on phase 2 parameter & 0.0017\\
		$\alpha_2$ & Approximation of $x$ on phase 2 parameter & 1.07Hz\\
		$\beta_2$ & Approximation of $y$ on phase 2 parameter & 0.34Hz\\
		\hline	
	\end{tabular} 	
\end{center}\captionof{table}{Approximation parameters}\label{approxParam}
\subsection{Simplified formulas \eqref{t1simple} and \eqref{LamdaTildSimple}} \label{intParamSimple}
We give here the simplified formulas of the intermediary parameters used to obtain \eqref{t1simple} and \eqref{LamdaTildSimple}.
\paragraph{Phase 1 parameters}
In this phase $H_1 \gg 1$, yielding
\beq
\begin{array}{r c l}
	\alpha_1 & \approx & 1.1 + KH_1 \approx KH_1,\\
	A_1 & \approx & -0.9, B_1 \approx 1, \beta_1 \approx LH_1 \text{ and } C_1 \approx \exp\left(0.9\cfrac{L}{K}\right).
\end{array}
\eeq
Using these values we can compute $\Gamma_1 - \Gamma_9$:
\beq \label{gammasSimple}
\begin{array}{r c l c l}
\Gamma_1 & \approx & \cfrac{J-2.9LH_1}{2.9JLH_1} & \sim & 0.24 \textit{ (with our parameters)}\\
\Gamma_2\ln\left(\cfrac{1}{1+\Gamma_3}\right) & \approx & \cfrac{1}{2.9LKH_1^2}\ln\left(\cfrac{1}{1-0.9\cfrac{L}{K}}\right) & \sim & 10^{-5}\\
\Gamma_4 & \approx & -\cfrac{0.9}{2.9LH_1} & \sim & 10^{-3} \\
\Gamma_5 & \approx & -\exp\left(-0.9\cfrac{L}{K}\right) & \sim & 0.5 \\
\Gamma_6 & \approx & 0.9\exp\left(-0.9\cfrac{L}{K}\right) & \sim & 0.46 \\
\Gamma_7 & \approx & \cfrac{0.81}{2.9LH_1}\left(\cfrac{LH_1}{LH_1}-1\right) & \sim & 10^{-6}\\
\Gamma_8 & \approx & \cfrac{0.81}{8.41LH_1} & \sim & 10^{-4}\\
\Gamma_9 & -\approx & \cfrac{0.81}{8.41LH_1} & \sim & 10^{-4}\\
\end{array}
\eeq
In our parameter range, we neglect the terms $\Gamma_2\ln\left(\cfrac{1}{1+\Gamma_3}\right)$, $\Gamma_4$, $\Gamma_7$, $\Gamma_8$ and $\Gamma_9$ in formula \eqref{t1}.
\paragraph{Phase 2 parameters}
In this phase, $|H_2|= O(1)$ and thus we can neglect the second order terms in $K$ and $L$, that are small in our parameter range. It yields
\beq
\begin{array}{r c l}
	\alpha_2 & \approx & 1.1 + KH_2,\,\, x(t_1) \approx 1 \text{ thus }
	A_2 \approx \cfrac{1}{1.1+KH_2}\exp\left((1.1+KH_2)t_1\right)\\
	& &\\
	B_2 & \approx & \cfrac{0.1+KH_2}{1.1+KH_2}, \,\, \beta_2 \approx \cfrac{0.37+0.34KH_2+0.1LH_2}{1.1+KH_2} \text{ and } y(t_1) \approx 0 \text{ thus } C_2 \approx 0.
\end{array}
\eeq
We can use these formulas to simplify the formulas for $\Lambda_1 - \Lambda_9$:
\beq \label{lambdasSimple}
\arraycolsep=1.4pt\def\arraystretch{2.5}
\begin{array}{r c l}
	\Lambda_1 & \approx &\cfrac{J(0.1+KH_2)-1.1-KH_2-0.29LH_2}{(1.1+KH_2+0.29LH_2)J}, \\
	\Lambda_2 & \approx & -\cfrac{1}{3.19LH_2}, \\
	\Lambda_3 & \approx & \cfrac{LH_2}{0.1LH_2+0.34KH_2+0.37}\\
	\Lambda_4 & \approx & \cfrac{LH_2}{29(0.37+0.34KH_2+0.1LH_2)}\\
	\Lambda_5 & \approx & \cfrac{1.1+1.2KH_2-0.037t_1-0.4KH_2t_1-0.11LH_2t_1}{1.18+2.26KH_2+0.32LH_2}\\
	\Lambda_6 & \approx & \cfrac{0.4+0.74KH_2-0.14t_1-0.26KH_2t_1-0.4LH_2t_1}{0.44+1.2KH_2+0.23LH_2}\\
	\Lambda_7 & \approx & \cfrac{-1.1LH_2+0.37LH_2t_1}{0.44+1.2KH_2+0.23LH_2}\\
	\Lambda_8 & \approx & \Lambda_9 \approx o(KH).
\end{array}
\eeq
\subsection{Termination time $t_2$ vs network connectivity J} \label{t2functionJ}
We fitted Equation \eqref{t2equation} defined with respect to the network connectivity parameter $J$ by a rational function
\beq
t_2(J) = \cfrac{AJ+B}{CJ+D}\left(1+\cfrac{\tau_{mAHP}}{J}\ln\left(\cfrac{h_{AHP}-T_{AHP}}{h(t_1)-T_{AHP}}\right)\right).
\eeq
\begin{figure}[h!] \centering
	\includegraphics[scale=0.8]{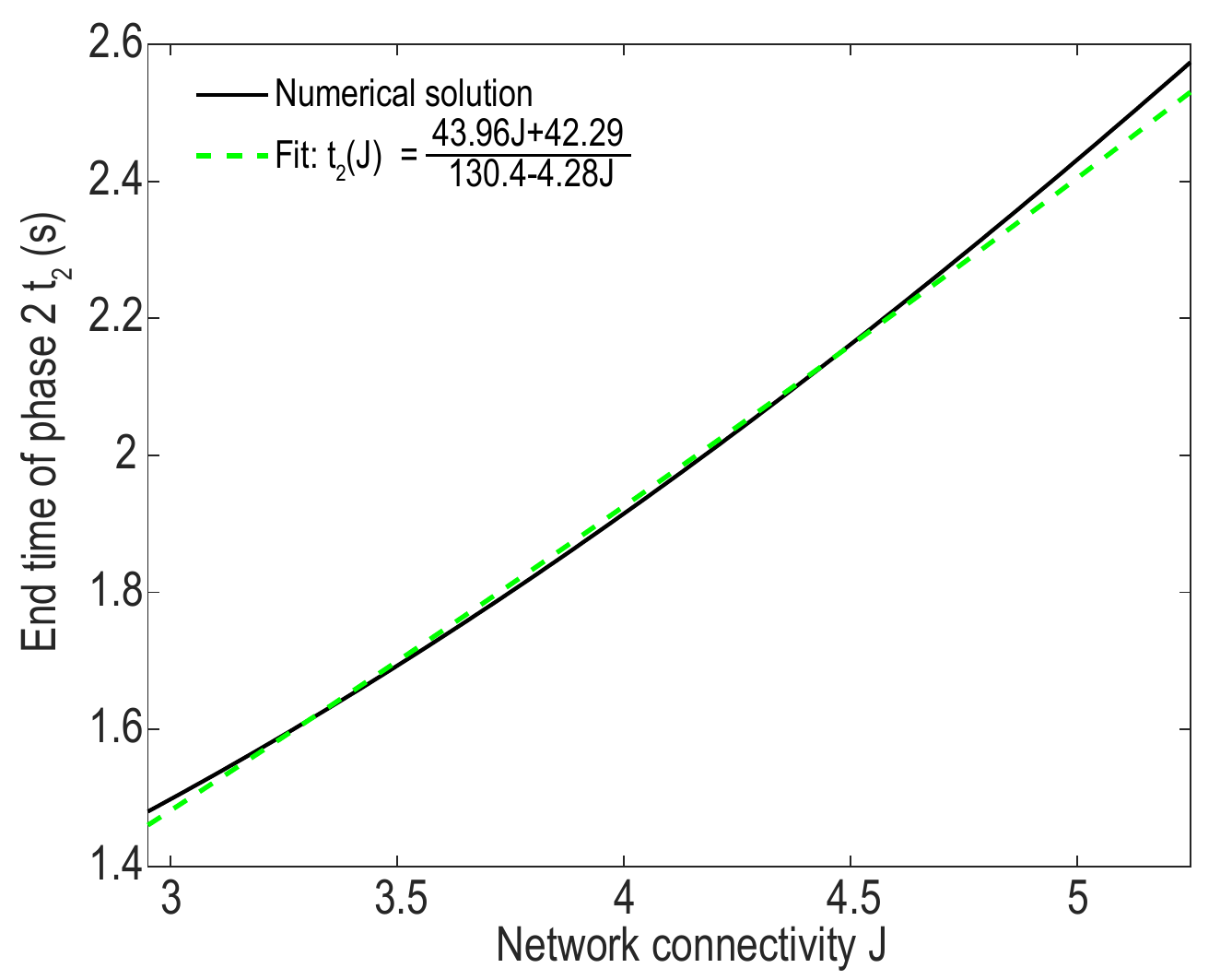}
	\caption{\textbf{Phase 2 termination time vs network connectivity J:} numerical solution \eqref{t2equation} (solid black) fitted by a rational function of $J$ (dashed green).} \label{t2functionJfigure}
\end{figure}

\newpage
\normalem
\bibliographystyle{IEEEtranlz}
\bibliography{biblioAHP}
\end{document}